\documentclass[prb,superscriptaddress,twocolumn,nofootinbib,floatfix,preprintnumbers]{revtex4-1}
\usepackage{epsfig,graphicx,tabularx}
\usepackage{amsmath,amsfonts,amssymb}
\usepackage{color,braket}
\usepackage{comment}

\usepackage{appendix}
\usepackage{dsfont}
\usepackage{amssymb}
\usepackage{bm}
\usepackage{ulem}
\usepackage{soul}
\usepackage{natbib}
\usepackage{nccmath}
\usepackage{array}
\newcommand{\beq}{\begin{equation}}
\newcommand{\eneq}{\end{equation}}
\newcommand{\be}{\begin{equation}}
\newcommand{\eeq}{\end{equation}}
\newcommand{\bea}{\begin{eqnarray}}
\newcommand{\eeqa}{\end{eqnarray}}
\usepackage{multirow}
\usepackage{wasysym}
\usepackage{hyperref}

\definecolor{forestgreen}{rgb}{0.4, 0.1, 0.7}
\definecolor{color2}{rgb}{0.1, 0.6, 0.1}

\hypersetup{
colorlinks=true,
citecolor=blue,
linkcolor=blue,
urlcolor=blue,
}



\begin{document}

\title{Bound state dynamics in the long-range spin-$\frac{1}{2}$ XXZ model}
\author{T. Macr\`i}
\email{macri@fisica.ufrn.br}
\affiliation{Departamento de F\'i­sica Te\'orica e Experimental,Universidade Federal do Rio Grande do Norte, 59072-970 Natal-RN, Brazil.}
\affiliation{International Institute of Physics, Universidade Federal do Rio Grande do Norte, 59078-400 Natal-RN, Brazil.}

\author{L. Lepori}
\affiliation{Dipartimento di Fisica, Universit\'a della Calabria, Arcavacata di Rende I-87036, Cosenza, Italy.}
\affiliation{I.N.F.N., Gruppo collegato di Cosenza, Arcavacata di Rende I-87036, Cosenza, Italy.}

\author{G. Pagano}{\color{black}
\affiliation{Department of Physics and Astronomy, Rice University,\\ 6100 Main Street, Houston, TX 77005, USA.}

\author{M. Lewenstein}
\affiliation{ICFO - Institut de Ci\`encies Fot\`oniques, The Barcelona Institute of Science and Technology, 08860 Castelldefels (Barcelona), Spain.}
\affiliation{ICREA, Pg. Lluis Companys 23, ES-08010 Barcelona, Spain.}

\author{L. Barbiero}
\email{luca.barbiero@polito.it}
\affiliation{Institute for Condensed Matter Physics and Complex Systems, DISAT, Politecnico di Torino, I-10129 Torino, Italy.}
\affiliation{ICFO - Institut de Ci\`encies Fot\`oniques, The Barcelona Institute of Science and Technology, 08860 Castelldefels (Barcelona), Spain.}

\begin{abstract}
Experimental platforms based on trapped ions, cold molecules, and Rydberg atoms have made possible the investigation of highly-nonlocal spin-${1/2}$ Hamiltonians with long-range couplings. Here, we study the effects of such non-local couplings in the long-range spin-${1/2}$ XXZ Heisenberg Hamiltonian. We calculate explicitly the two-spin energy spectrum, which describes all possible energetic configurations of two spins pointing in a specific direction embedded in a background of spins with opposite orientation. For fast decay of the spin-spin couplings, we find that the two-spin energy spectrum is characterized by well-defined discrete values, corresponding to bound states, separated by a set of continuum states describing the scattering region. In the deep long-range regime instead, the bound states disappear as they get incorporated by the scattering region. The presence of two-spin bound states results to be crucial to determine both two- and many-spin dynamics. On one hand, radically different two-spin spreadings can be observed by tuning the decay of the spin couplings. On the other hand, two-spin bound states enable the dynamical stabilization of effective antiferromagnetic states in the presence of ferromagnetic couplings. Finally, we propose a novel scheme based on a trapped-ion quantum simulator to experimentally realize the long-range XXZ model and to study its out-of-equilibrium properties.
\end{abstract}

\maketitle

\section{Introduction}
The investigation of quantum many-body
systems driven in out-of-equilibrium congurations represents
an intriguing topic of current research \cite{silva,rigol,gogolin,dalessio,essler2}. For instance, dynamical procedures based on Floquet time periodic modulations \cite{goldman,bukov,eckardt} in interacting systems have allowed for the realization of lattice gauge theories \cite{barbiero,tilman,monika} and Hall-like states of matter \cite{greiner}. Moreover, schemes making use of adiabatic evolutions have been proved to be particularly relevant for quantum computation realizations \cite{das,santoro,albash} and the study of the Kibble-Zurek mechanism \cite{lukin}. A further example of a possible out-of-equilibrium evolution is represented by sudden quenches, where an initial state is time-evolved with a specific time-independent target Hamiltonian. In the case of energy-conserving evolution, quantum quenches have revealed several interesting many-body quantum effects ranging from entanglement propagation \cite{calabrese}, many-body localization \cite{bloch} and long-range hopping \cite{nagerl}, to dynamical string breaking \cite{pichler}, entangled edge states \cite{barbiero2018} and Hilbert space fragmentation \cite{monika2}. Interestingly, some dynamical effects can also be predicted by looking at the peculiar two-body energy spectrum. The latter describes all the possible energetic configurations of two interacting particles and is usually characterized by a scattering region with extended wave-functions, and by two-body bound states with localized wave-functions \cite{RBP1}. In this respect, it has been shown that the dynamical occupation of bound states enables the investigation of bound-pairs with slow dynamical evolutions \cite{RBP1,RBP2}. Crucially, also pure many-body effects, such as disorderless quasi-many-body-localization \cite{io,li2020}, quantum magnetism \cite{muth}, and metastable Mott states \cite{nagerl2} can be induced entirely by the presence of two-body bound states.

A paradigmatic model, where most of the aforementioned effects can be studied is represented by the short-range integrable spin-$1/2$ XXZ Heisenberg Hamiltonian. This model is currently available in experiments involving bosonic mixtures \cite{fukuhara,ketterle} and it represents a fundamental tool to explore quantum magnetism in low dimensional materials \cite{book_magnetism, giuliano2013}. Moreover, it has been shown that fundamental results concerning magnon propagation \cite{fukuhara,essler,io3} and magnetic ordering \cite{altman,io2} can be interpreted as the consequence of the bound states characterizing the two-spin energy spectrum of the integrable XXZ model. The physics of the XXZ Hamiltonian results to be much less investigated in presence of long-range couplings breaking the model integrability. As indeed known, long-range couplings can drastically increase the level of complexity of quantum systems \cite{dauxois,kestner1,CDP2,CDP3,hermes2020,hauke2013,maghrebi,LRB,LRB1,LRB2,koffel2012, AL1,AL2,kastner2014,vodola2014,gorshkov2016,vodola2016,daley2016,lepori2017,lepori2018,pezze2018,lepori2016,Defenu2021,gori2013,leporiLR}.

Crucially, systems with long-range couplings are currently available in different experimental platforms working at ultracold temperatures ranging from dipolar gases \cite{pfau,Bohn2017} and atoms in cavities \cite{risch}, to Rydberg atoms \cite{schauss2015,labuhn2016,browaeys2020,morgado2021}, and trapped ions \cite{ions}. While in most of the aforementioned examples the decay of long-range spin-spin couplings is fixed, trapped-ion systems offer the possibility to tune such quantity \cite{Britton2012}. This feature has allowed to design experiments reproducing spin-$1/2$ long-range Hamiltonians investigating crucial aspects of lattice gauge theories \cite{LGT1}, time crystals \cite{TC,Kyprianidis2021}, many-body-localization \cite{MBLi,morong2021observation}, dynamical phase transitions \cite{DPTm,DPTb}, correlation propagation \cite{Es,richerme2014,tan2021domain} and many-body dephasing \cite{Kaplan2020}.

Motivated by such recent and flourishing activity, in this work we investigate the static and dynamical properties of the the long-range XXZ Hamiltonian. In section \ref{model}, we introduce the model and discuss how the long-range character of the spin-spin couplings modifies the single-spin bandwidth and, consequently, the two-spin energy spectrum, introduced in section \ref{2spinex}. This quantity describes all the possible configurations of a system composed by two spins pointing in a specific direction and all the other spins with an opposite orientation. In this work we show that the two-spin energy spectrum is strongly affected by different choices of the Hamiltonian parameters: On one hand, our calculations reveal the presence of well-defined bound states with localized wave functions in the case of fast decay of the spin-spin couplings and relatively large ferromagnetic interaction along the $z$ direction. These discrete solutions turn out to be sensibly separated by a scattering region where the wave function is totally delocalized. On the other hand, when the Hamiltonian couplings are slowly decaying, the model phenomenology is dramatically different. In this scenario, our results show that the two-spin bound states get incorporated by the scattering region and, as a consequence, the particle wave function delocalizes for any finite coupling strength. These two distinct regimes support very different out-of-equilibrium properties. In particular, as shown in section \ref{2spindynamics}, our exact calculations clarify that both slow and fast expansions can be induced by adjusting the decay of the couplings between spins. In section \ref{Manybodydynamics}, we show that two-spin bound states can have deep consequences even in many-spin configurations, as for the case of vanishing total magnetization. Indeed, for appropriate Hamiltonian parameters, we show that the presence of bound states can gives rise to an effective two-spin antiferromagnetic blockade. The latter, consisting of the partial conservation of the number of antiferromagnetic domains, is able to produce intriguing magnetic states. More precisely, our calculations show that, for a specific state preparation, an antiferromagnetic state in presence of ferromagnetic couplings can be dynamically
stabilized. Finally, in section \ref{trappedion}, we present a discussion explaining how the long range XXZ model and its different dynamical regimes can be achieved and detected with a trapped-ion quantum simulator.  
\section{The model}
\label{model}
The long-range XXZ model is described by the following Hamiltonian
\begin{equation}
H=\sum_{i<j}\frac{J}{|i-j|^{\alpha}}\left(S^{+}_iS^{-}_j+S^{-}_iS^{+}_j+2\,\Delta S^{z}_iS^{z}_j \right) \, ,
\label{hamspin}
\end{equation}
where $S_{i}^{\pm}=S_{i}^x \pm \imath S_{i}^y$, and $S_i^{\gamma}$ with $\gamma=x,y,z$ are standard spin-$1/2$ operators describing a system of $L$ spins coupled by $J=-1$ (fixing our energy scale) along the $x,y$ directions and by $2\,J\,\Delta$ along the the $z$ direction. 
Crucially, here we consider the couplings along the three directions decaying as function of the inter-spin distance $|i-j|^\alpha$. The long-range model in Eq. (\ref{hamspin}) has been recently realized both in an experiment exploring the motional sidebands in a trapped ensemble of ultracold bosonic atoms \cite{rey} and, with fixed $\alpha=3$, in Rydberg atomic platforms \cite{geyer,scholl}. Moreover, proposals to explore the same $\alpha=3$ regime in polar molecules experiments are available \cite{molecules}. The possibility to tune $\alpha$ as in trapped-ion experiments, is particularly important since this term can drive the system from an effective short-range regime for $\alpha>1$ to the highly non-local case $\alpha\leq1$. In this context, for vanishing total magnetization $\sum_{i=1}^LS^z_{i}=0$, it has been demonstrated that a large $\alpha$ generates the physics of the short range XXZ integrable model \cite{maghrebi,frerot}.  On the other hand, when $\alpha\leq1$, the ground state can be characterized by a spontaneous continuous symmetry breaking \cite{maghrebi}. As indirectly predicted by the Mermin-Wagner theorem \cite{mermin}, which holds only for local Hamiltonians, this last effect results to be a direct consequence of the pure long-range couplings. 
The role of the exponent $\alpha$ turns out to be crucial also at a single-spin level, namely when a single spin up is embedded in a distribution of spins pointing down, or viceversa, {e. g.} $\sum_{i=1}^LS^z_{i}=|L-2|/2$ and $\Delta=0$.
In particular, for this specific configuration we find that Eq. (\ref{hamspin}) has a single-spin spectrum in momentum space $\lambda_{\alpha} (k) = J\, g_{\alpha} (k)$, with $g_{\alpha} (k) =  \sum_{l =1}^{L} \, \frac{ \cos k l}{l^{\alpha}}$ and $0 \leq k < 2 \pi$, in steps of $\frac{2 \pi}{L}$. Notice that the functions $g_{\alpha} (k)$ can be evaluated in the thermodynamic limit, in terms odd polylogarithmic functions: $g_{\alpha} (k) = \mathrm{Im} \, \mathrm{Li}_{\alpha} (e^{ik})$ \cite{grad,abram,olver}. Relevantly, it turns out that $g_{\alpha} (k)$ keeps basically the same value for $\alpha>1$. On the other hand, the energy of the single-spin spectrum starts to increase drastically when $\alpha$ approaches $1$, where $g_{\alpha} (k)$ diverges at $k = 0,\, \pi$. As a consequence, in this genuine long-range regime the relative bandwidth $W_{\alpha}^{(s)} = 2 J g_{\alpha} (k=0)$ increases indefinitely. Crucially, $W_{\alpha}^{(s)}$ represents a very important quantity when dealing with the dynamics of isolated quantum systems. Indeed, $2 \, W_{\alpha}^{(s)} $ is the effective kinetic energy that particles can dynamically convert into potential energy while keeping the condition of energy conservation fulfilled. Analogously, in terms of spins, $2 \, W_{\alpha}^{(s)} $ represents the amount of energy that the system can dynamically convert from magnetically coupled spins along the $x,y$ directions to form magnetic domains along the $z$ direction. As a consequence, the peculiar behavior of $W_{\alpha}^{(s)}$ suggests that for $\alpha\to1$ the amount of convertible energy is unbounded. Since this last scenario is not compatible with presence of bound states, in the next section we calculate the two-spin energy spectrum corresponding to the case $\sum_{i=1}^LS^z_{i}=|L-4|/2$ at finite $\Delta$.
\begin{figure*}[t]
\centering
 \includegraphics[width=.995\linewidth]{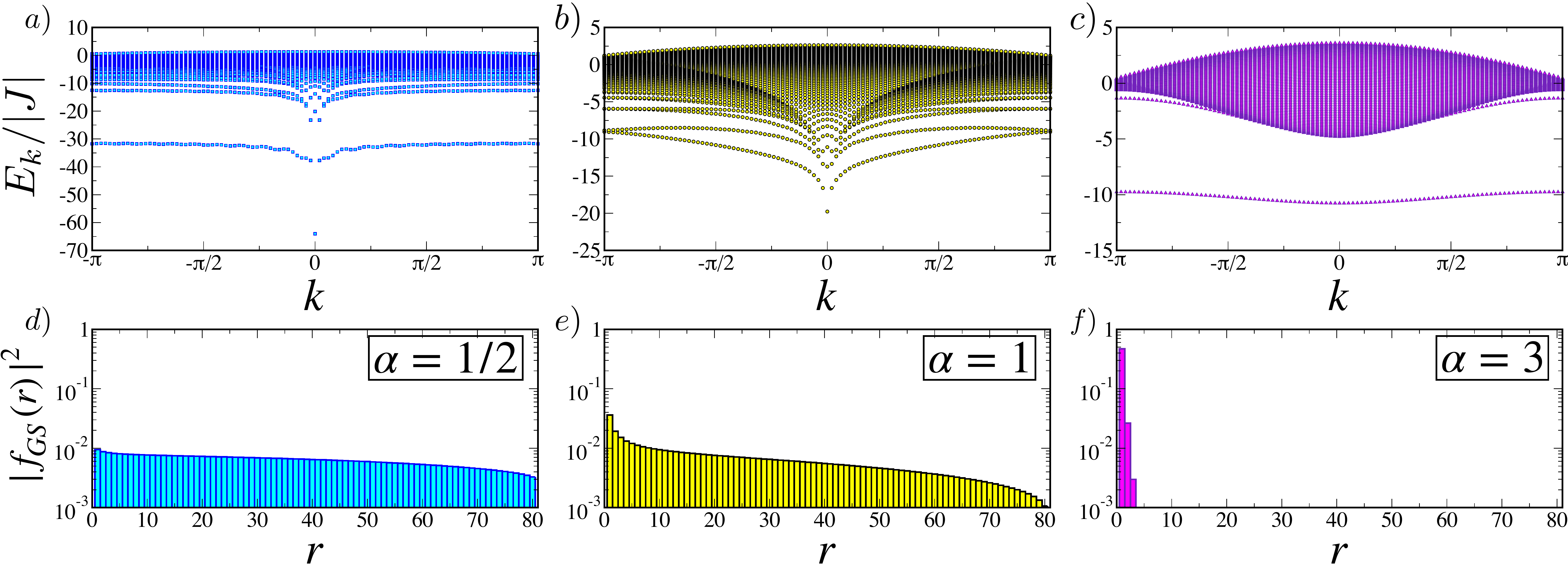}
\caption{
\textit{Upper panels}: Energy spectrum of the two-spin configuration $\sum_{i=1}^LS^z_{i}=|L-4|/2$, $L = 100$,
as a function of the momentum $k\, L \in [-\pi,\pi ]$ for $\alpha=1/2$ \textit{a)}, $\alpha=1$ \textit{b)} and $\alpha=3$ \textit{c)}. \textit{Lower panels}: 
density $|f_{\it GS} (r)|^2$, i.e. the lowest energy eigenstate corresponding to $k=0$ (see main text), for $\alpha=1/2$ \textit{d)}, $\alpha=1$ \textit{e)}, and $\alpha=3$ \textit{f)}.
Here we consider ferromagnetic couplings with $\Delta/J=-5$ and solve eq.(\ref{eqBS})
for $p_{max}=80$ (half-maximum relative distance).
For $\alpha\le 1$ the relative wavefunction is non vanishing even at the edges.
This feature persists even for larger values of $\Delta$.}
\label{fig:one}
\end{figure*}

\section{Two-spin excitation spectrum}
\label{2spinex}
The two-spin energy spectrum relative to Eq. \eqref{hamspin} can be derived by calculating explicitly the two-spin wave function equation in the frame of the center-of-mass. In particular, we consider a system with $\Delta/J<0$, containing two spins pointing up and all the other spins pointing down. The relative wave function can then be written as:
\begin{equation}
\left|\psi_{2S}\right> = \sum_{i<j}\psi(i,j) S_{i}^+ S_{j}^+ \prod_{i=1}^L 
\left|\downarrow \right>.
\end{equation}
By changing variables and moving to the relative coordinates of the center-of-mass $r=|i-j|$ and $R=|i + j|/2$  one can rewrite the previous equation as
\begin{equation}
\left|\psi_{2S}\right> = \sum_{2R,r=-\infty}^\infty\psi(R,r) S_{R+\frac{r}{2}}^+ S_{R-\frac{r}{2}}^+ \prod_{i=1}^L 
\left|\downarrow \right>.
\end{equation}
Only symmetric states with respect to particle exchange ($r\rightarrow -r$) are part of the Hilbert space of the model with only two symmetric excitations. 
We first consider the action of the diagonal Hamiltonian 
$H_d=2\,J\Delta\sum_{i<j}\frac{S^{z}_iS^{z}_j}{|i-j|^{\alpha}}$ on the states with zero and only one $\ket{\uparrow}$ spin respectively 
\begin{eqnarray}
 H_d \prod_i^L \ket{\downarrow}_i &&=2\,J\Delta \sum_{i<j} \frac{1}{|i-j|^\alpha} \frac{1}{4}  \prod_{i=1}^L \ket{\downarrow}_i \nonumber \\
&& =2\,J\Delta\frac{L}{4}\zeta(\alpha)  \prod_{i=1}^L \ket{\downarrow}_i
\end{eqnarray}
 \begin{equation}
 H_d\, S_R^+ \prod_{i=1}^L \ket{\downarrow}_i=2\,J \Delta \left(\frac{L}{4}-1 \right)
\zeta(\alpha) S_R^+ \prod_{i=1}^L \ket{\downarrow}_i 
\end{equation}
with $\zeta(\alpha)$ being the Riemann zeta. 
\begin{widetext}
Then $ H_d$ acts on the two $\ket{\uparrow}$ spins state as:
\begin{equation}
\begin{array}{ccl}
 H_d\, S_{R+\frac{r}{2}}^+ S_{R-\frac{r}{2}}^+ \prod_{i=1}^L  
\left|\downarrow \right>_i &=& 2\,J \Delta \left[(\frac{L}{4}-2) \zeta(\alpha) +\frac{1}{r^\alpha} \right]
S_{R+\frac{r}{2}}^+ S_{R-\frac{r}{2}}^+ \prod_{i=1}^L  \ket{\downarrow}_i 
\end{array}
\end{equation}
The off-diagonal Hamiltonian term, 
$H_o=\sum_{i<j}\frac{J}{|i-j|^{\alpha}}\left(S^{+}_iS^{-}_j+S^{-}_iS^{+}_j \right)$, acts on the two-spin state as:
\begin{equation}
\begin{array}{lll}
H_o\, S_{R+\frac{r}{2}}^+ S_{R-\frac{r}{2}}^+ \prod_{i=1}^L  
\left|\downarrow \right>_i =&& \\ \\ 
J \sum_{p=1}^\infty \frac{1}{p^\alpha} \left( 
S_{R-\frac{r}{2}-p}^+ S_{R+\frac{r}{2}}^+ +
S_{R-\frac{r}{2}+p}^+ S_{R+\frac{r}{2}}^+ +
S_{R+\frac{r}{2}-p}^+ S_{R-\frac{r}{2}}^+ +
S_{R+\frac{r}{2}+p}^+ S_{R-\frac{r}{2}}^+
\right) \prod_{i=1}^L  \left|\downarrow \right>_i 
\end{array}
\end{equation}
We now take the following product form of the coefficients of the wavefunction:
\begin{equation}
\psi_{2S}(R,r) = e^{i k R} f(r).
\end{equation}
Substituting into the eigenvalue equation $ H \left|\psi_{2S}\right>=E\left| \psi_{2S}\right>$ we get:
\begin{equation}
\begin{array}{lcl}
J \sum_{2R,r=-\infty}^\infty e^{i k R} 
 \sum_{p=1}^\infty \frac{1}{p^\alpha}
\left( 
e^{-i k \frac{p}{2}}f(r+p) + e^{i k \frac{p}{2}}f(r-p) + e^{-i k \frac{p}{2}}f(r-p) + e^{i k \frac{p}{2}}f(r+p)
\right) +&& \\ \\
2\,\Delta f(r) e^{i k R} \left[\left(\frac{L}{4}-2 \right) \zeta(\alpha)+\frac{1}{r^\alpha} \right]
S_{R+\frac{r}{2}}^+ S_{R-\frac{r}{2}}^+ \prod_{i=1}^L  \left|\downarrow \right>_i
= 
&&\\ \\
E \sum_{2R,r=-\infty}^\infty 
e^{i k R}f(r)S_{R+\frac{r}{2}}^+ S_{R-\frac{r}{2}}^+ \prod_{i=1}^L  \left|\downarrow \right>_i.
\end{array}
\end{equation}
Therefore the equation for the coefficients $f(r)$ reads:
\begin{equation}
2\,J \sum_{p=1}^\infty \frac{1}{p^\alpha} \cos\left( \frac{k \, p}{2}\right)\left(f(r+p) +f(r-p) \right) +\frac{2\,J\,\Delta}{r^\alpha}f(r) = E_k\, f(r),
\label{eqBS}
\end{equation}
\end{widetext}
where we subtracted the constant energy $\propto (\frac{L}{2}-2)\zeta(\alpha)$.
Eq.(\ref{eqBS}) is the equation we solve numerically to extract the two-spin energy spectra. 
In particular, one can notice in Fig.\ref{fig:one} that different choices of $\alpha$ give rise to two-spin energy spectra with very distinct features.

 \begin{figure}[t!]
 \centering
\includegraphics[width=.995\linewidth]{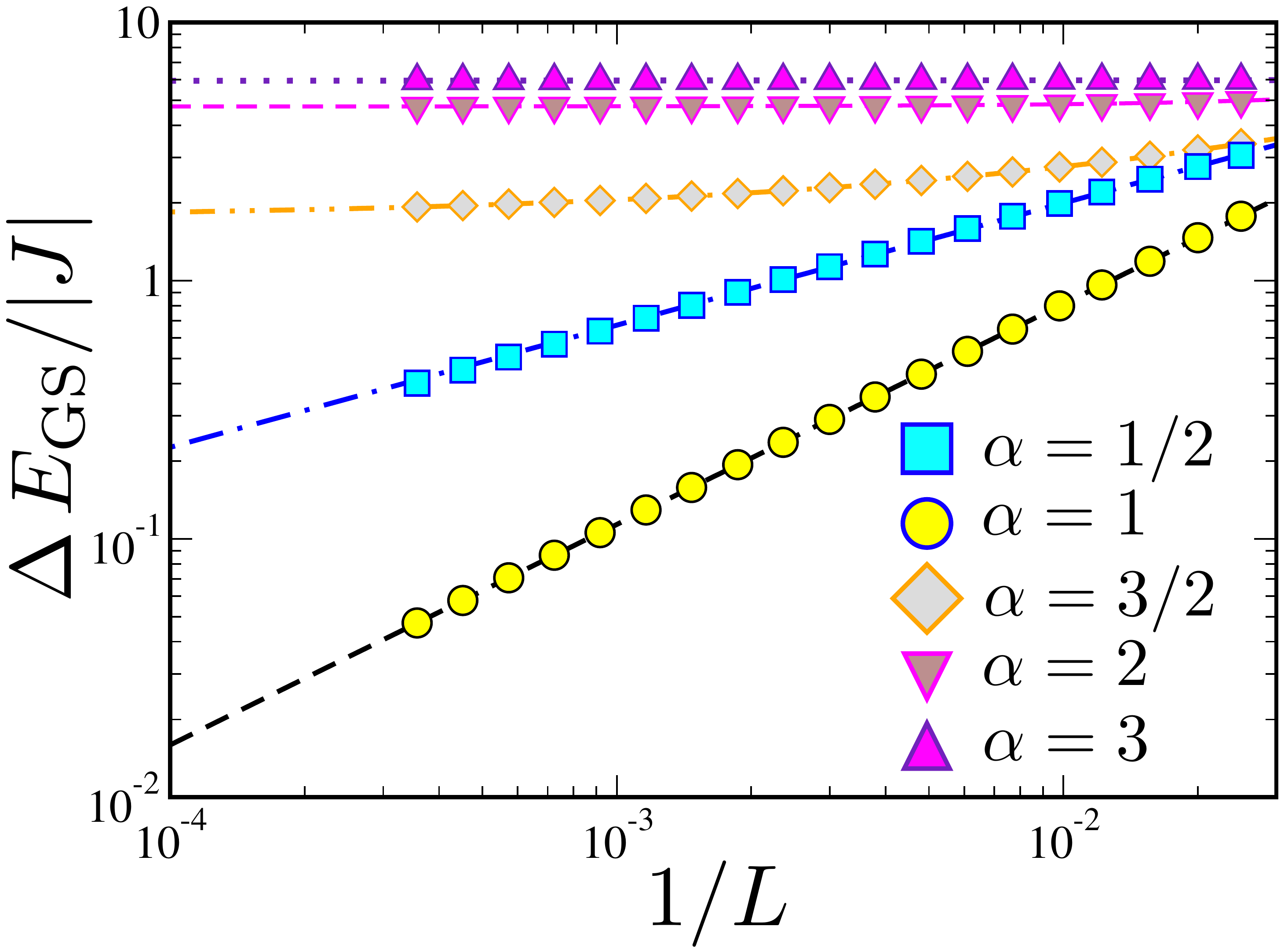}
\caption{Plot of the difference $\Delta E_\text{GS}=E_{GS}(\Delta=0)-E_{GS}(\Delta=-5)$ 
as a function of system size $L$ for different values of $\alpha$. 
Whereas for $\alpha>1$ the energy difference
$\Delta E_{GS}$ converges to the energy of the bound state relative to the scattering state at finite $\Delta$,
for $\alpha\le 1$ the ground state at $\Delta>0$ is {\it absorbed} by the 
scattering (continuum) states and $\Delta E_\text{GS} \rightarrow 0$ in the 
thermodynamic limit. 
Fitting functions with free parameters $\{a_0,a_1,a_2 \}$ for $\alpha \le 1$ are given by $\Delta E = \frac{a_0 }{L^{a_{1}}}$, 
whereas for $\alpha>1$: $\Delta E = \frac{a_0 }{L^{a_{1}}}+{a_2}$ 
with $a_2=1.75$ ($\alpha=3/2$), $a_2=4.72$ ($\alpha=2$), and $a_2=5.95$ ($\alpha=3$).
}
\label{fig:two}
\end{figure}

The upper panels in Fig. \ref{fig:one} shows the two-spin energy spectrum for $\alpha= 1/2, 1, 3$ and a ferromagnetic coupling in the $z$ direction $\Delta/J=-5$. The three reported cases are clearly characterized by a high density of possible states around the region of vanishing energies. This energetic sector represents a scattering band where the energy eigenstates are spread into the bulk and the wave function is fully delocalized. Furthermore, at $\alpha=3$ a clear well-defined energetic state occurs at lower energy. As visible by looking at the corresponding relative wave function reported in the lower panel of Fig. \ref{fig:one}, the latter corresponds to a bound state where two nearest neighbors spins pointing in the same direction are energetically bounded and form a new localized effective spin. Note that different values of $\Delta/J$ simply modify the effective separation existing between the bound state and the scattering region. It is also worth to underline that, in analogy with previous studies on a similar model \cite{io}, for $\alpha=3$ more negative values of $\Delta/J$ can also give rise to extended bound states, namely two localized spins pointing in the same direction with in between an arbitrary number of spins with opposite orientation. As also visible in Fig. \ref{fig:one}d-e, the presence of bound states at $k=0$ appears to be much less evident for $\alpha=1, 1/2$. Here the low-energy lines merge with the region of scattering states for any ratio $\Delta/J$. In order to clarify this point, the shape of the two-spin wave function $|f_{\it GS} (r)|^2$, i.e. the lowest energy eigenstate corresponding to $k=0$, can reveal interesting aspects. In particular, we show that, contrary to the $\alpha=3$ case, for $\alpha=1, 1/2$ the relative wave function is delocalized and clearly non vanishing even at the edges. Moreover, this feature persists for smaller values of $\Delta/J$ and system sizes $L$. As a consequence, one expects that the bound states get incorporated into the scattering region and no localized solutions exist. In order to enforce this conclusion, in Fig. \ref{fig:two} we plot the size dependence of the energy difference between the interacting ($\Delta/J=-5$) and the non-interacting case ($\Delta/J=0$) at vanishing momentum $k$, namely $\Delta E_{GS}=E_{GS}(\Delta/J=0)-E_{GS}(\Delta/J=-5)$. While for $\alpha>1$ the energy difference $\Delta E_{GS}$ converges to the energy of the bound state relative to the lowest scattering (continuum) state for the corresponding value of $\Delta/J$, for $\alpha\le 1$ the ground state is {\it absorbed} into the 
scattering states and $\Delta E_{GS} \rightarrow 0$ in the thermodynamic limit.
\begin{figure*}[t]
\centering
\includegraphics[width=0.995\linewidth]{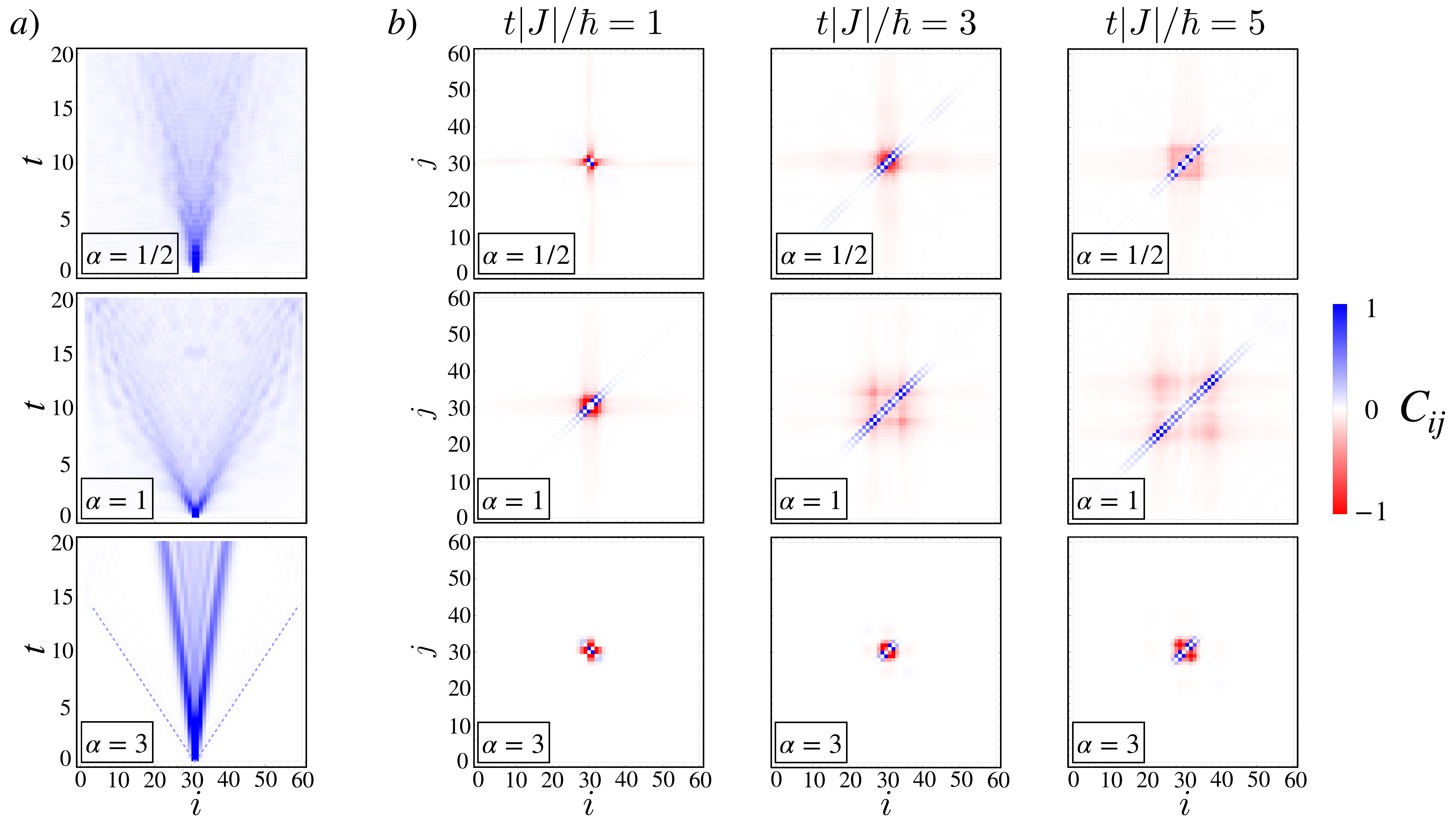}
\caption{Dynamics for two initially localized excitations at the sites $i=30$ and $j=31$, 
on a chain with length $L=60$ and open boundary conditions,  for $\alpha=1/2$ (top row), $\alpha = 1$ (center row), and $\alpha = 3$ (bottom row), and $\Delta/J=-5$.
Left \big(a)\big) column: time evolution of the expectation value $\langle P_i \rangle$ of the operator
$P_i= S^z_i(t)+1/2$, measuring the fraction of spins in the $\uparrow$ configurations, as a function of time. 
Panels b): correlation functions, $C_{ij}= \langle P_iP_j\rangle- \langle P_i\rangle \langle P_j \rangle$ between the positions of two excitations, for times $t|J|/\hbar=1$ (left column), $t|J|/\hbar=3$ (central column), and $t|J|/\hbar=5$ (right column).
Color scales are normalized for each image to its peak value.}
\label{fig:tre}
\end{figure*}
The absence of bound states turns out to be a direct consequence of the unbounded bandwidth $W(k)$ peculiar of the $\alpha\to1$ case. Here, the system can convert an almost unbounded amount of kinetic energy into potential energy while keeping the total system energy conserved. This is not the case when $\alpha$ is larger and long-range effects are less relevant. More precisely, in this case the finite bandwidth implies that for $\Delta/J$ much smaller than zero, a ferromagnetic domain formed by two spins cannot be dynamical destroyed without violating the energy conservation. These properties result to be very relevant when exploring out-of equilibrium configurations both at two- and at many-spin level.

\section{two-spin dynamics}
\label{2spindynamics}
In order to clarify how the peculiar shape of the two-spin spectrum affects the out-of-equilibrium properties of the system, in this section we study the evolution of an initial product state $|\downarrow_1\downarrow_2...\downarrow_{L/2-1}\uparrow_{L/2}\uparrow_{L/2+1}\downarrow_{L/2+2}...\downarrow_{L-1}\downarrow_L\rangle$ which describes a state with total magnetization $\sum_{i=1}^LS^z_{i}=-(L-4)/2$ and two antiferromagnetic domains given by the different orientation of the two central spins. More precisely, by means of exact diagonalization calculations, we evolve  this state through the Hamiltonian in Eq. (\ref{hamspin}), with ferromagnetic coupling $\Delta/J=-5$. In the first column of Fig. \ref{fig:tre}, we report the time evolution of the expectation value $\langle P_i \rangle$ of the operator $P_i= S^z_i(t)+1/2$, measuring the fraction of spins in the $\ket{\uparrow}$ state. By tuning $\alpha$, one can explore very different expansion dynamics. In particular, for $\alpha=3$ we show the presence of two main signals. The first one is associated with the slow expansion of the bound state which gives rise to the peculiar cone-like propagation, as predicted by the Lieb-Robinson bound \cite{LRB1}. More precisely, the initial state has two antiferromagnetic domains and this number can not  be sensibly changed without violating the requirement of energy conservation. 
 As a consequence, the two central spins form a bound state propagating along the chain with an effective second order hopping processes $\tilde{t}^{(D)}_{ij}$ (see the appendix for details) much slower than the single-spin spreading $J$. 
In the same lower panel of Fig. \ref{fig:tre} a), a much weaker signal is also present, which is associated to the expansion of residual unbound spins, in analogy with the short-range case \cite{essler,fukuhara}. Notice that we expect a residual weak single spin spreading signal as long as $\Delta/J$ is finite.  As also clear in Fig. \ref{fig:tre}a, different choices of $\alpha$ strongly modify the aforementioned picture. Indeed, it is possible to see that for $\alpha=1/2,1$ a very fast spin propagation occurs, as expected for purely long-range systems \cite{richerme2014, Es}. In particular, we notice that the two spins remain totally unbound, as they rapidly expand, bouncing against the system edges and producing interference patterns in the center of the chain. As specified, this peculiar expansion is a direct consequence of both long-range couplings and the absence of bound states \cite{LRB2,richerme2014,Es}. In order to further clarify the very rapid expansion of unbound spins, in the right nine panels of Fig. \ref{fig:tre} we calculate the time dependent behavior of the connected two-body correlation function:
\begin{equation}
C_{ij}(t)=\langle P_iP_j\rangle- \langle P_i\rangle \langle P_j \rangle.
\label{corr}
\end{equation}
Here $\langle P_iP_j\rangle$ is the equal time correlation function which gives a measure of the time dependent probability of finding two $\uparrow$ spins at a certain distance $|i-j|$. In other words, Eq. (\ref{corr}) provides informations about the time that a spin at position $i$ needs to become correlated with another spin at position $j$. The behavior of $C_{ij}(t)$ clearly confirms the interpretation of our results. In particular, for $\alpha\leq1$ the two $\uparrow$ spins which at $t=0$ were in the center of the system get correlated at very large $|i-j|$ almost instantaneously, thus confirming that the single spin expansion velocity is almost unbounded. As we discussed, this effect is peculiar of long-range interacting systems where the usual Lieb-Robinson bound does not apply. On the other hand, the $\alpha=3$ case, which resembles the case of short-range effective coupling, displays a very different behavior of $C_{ij}(t)$. Indeed here, the two $\uparrow$ spins need a large time before becoming correlated event at intermediate distances thus reflecting the very slow spreading of the bound state. As also shown in a similar model \cite{io}, in the $\alpha=3$ case a different initial state can give rise to a very peculiar dynamics. In particular, if one considers an initial configuration where two $\uparrow$ spins are initially localized at large $|i-j|$ and let them evolve with Eq. (\ref{hamspin}) and ferromagnetic coupling $\Delta/J<0$ below a critical value, a very peculiar dynamics can take place. Here, due the sign of $\Delta/J$, the system would tend to minimize the number of antiferromagnetic domains, bringing them from 4 to 2, which would then correspond to the configuration where the two $\uparrow$ spins are at distance $|i-j|=1$. The crucial point is that, if $\Delta/J$ is negative enough, the state where the two spins sit next to each other is dynamically prohibited due to the energy conservation. As a consequence, an effective antiferromagnetic blockade is felt by the spins. This effect allows the $\uparrow$ spins to dynamically decrease their distance only up to a critical value $|i-j|_c$ which fixes the range of the antiferromagnetic blockade, namely the minimal distance that two $\uparrow$ spins can reach while keeping the total energy conserved. It is also worth to specify that $|i-j|_c$ is not fixed, since it depends by $\Delta/J$: for smaller values of $\Delta/J$ a larger $|i-j|_c$ can be achieved. This effective antiferromagnetic blockade is the analogous of the effective hard-core repulsion felt by  suddenly quenched bosons in the regime of attractive contact interaction. In this context, interesting many-body effects, like the presence of metastable Mott states with attractive interactions, have been experimentally achieved \cite{nagerl2}. For this reason, it turns to be of great relevance to understand how the peculiar shapes of the two-spin energy spectrum can affect the many-body dynamics of the model of Eq. (\ref{hamspin}).

\section{Many-spin dynamics}
\label{Manybodydynamics}

The antiferromagnetic blockade and the formation of two-spin bound state occurring in the $\alpha>1$ case have deep consequences even on the many-spin dynamics. Indeed, specific state preparations may allow for the investigation of interesting phenomena like disorderless quasi many-body localization and single-spin gluing \cite{io,li2020}. As an example, if one considers an initial state where both spins sitting next to each others and isolated spins are present and then time evolve this state with Eq. (\ref{hamspin}) at strongly negative $\Delta/J$ and $\alpha>1$, both bound states and the antiferromagnetic blockade take place. Here, by performing perturbation theory and neglecting residual interacting terms, one can derive the effective model
\begin{eqnarray}
H^{(\mathrm{eff})}&&=\sum_{i<j} (\tilde{t}^{(D)}_{ij} \, \tilde{D}_i^\dagger \tilde{D}_j + \tilde{w}^{(sD)}_{ij} \, \tilde{S}_i^{+} \tilde{D}_j^{\dagger} \, \tilde{D}_i \tilde{S}^-_j+\nonumber\\
&&+\frac{J\tilde{S}^+_i\tilde{S}^-_j}{2|i-j|^{\alpha}}+\mathrm{H. c.} )
\label{Heff_1}
\end{eqnarray}
where $\tilde{D}_i^\dagger=S_i^+S_{i+1}^+(\tilde{D}_i=S_i^-S_{i+1}^-)$ and $\tilde{S}_i^{+}(\tilde{S}_i^{-})$ refer to the raising (lowering) of bound states and single spins respectively, both experiencing the antiferromagnetic blockade. The coefficients $\tilde{t}^{(D)}_{ij}$ and $\tilde{w}^{(sD)}_{ij}$ both depend on $\alpha$ and $\Delta$, and can be derived by performing perturbation theory up to the second order, see appendix \ref{app2}. This effective model in Eq. \eqref{Heff_1} shows deep analogies with the one derived in \cite{io}, and we expect that similar localization properties can also occur in the fully long-range model Eq.(\ref{hamspin}).\\ 
On the other hand, by studying the many-spin dynamics of Eq. (\ref{hamspin}) we unveil the presence of a further counterintuitive effect. In particular, in Fig.\ref{fig:four} we employ time-dependent density matrix renormalization group (t-DMRG) \cite{dmrg,dmrg2} calculations to study the evolution of a product antiferromagnetic state $|\psi_{AF}\rangle=|\!\!\downarrow_1\uparrow_2...\downarrow_{L/2}\uparrow_{L/2+1}...\downarrow_{L-1}\uparrow_L\rangle$ evolving with Eq.\,(\ref{hamspin}) at fixed $\Delta/J=-5 $. More in detail, we study the time dependent expectation value of the staggered magnetization density $M_s(t)=\sum_i (-1)^i \left< S^z_i (t)\right>/L$ which represents the order parameter of antiferromagnetic states. In the case of nearest-neighbour couplings, it has been shown for $\Delta/J\ge0$ \cite{altman} that $M_s(t)$ remains finite at very long time when $\Delta/J$ is large enough, thus signaling the conservation of the antiferromagnetic order during the time evolution. On the other hand, a low or vanishing $\Delta/J$ gives rise to $M_s(t)=0$ after short times. In Fig \ref{fig:four} we show that, despite the ferromagnetic coupling $\Delta/J=-5$, an exponent $\alpha=3$ allows to $M_s(t)$ to remain finite during the time evolution meaning that the antiferromagnetic order is partially conserved. This effect might sound counterintuitive since the system keeps the antiferromagnetic ordering of the initial state despite the effective ferromagnetic coupling. The reason for such a phenomenon mainly lies in the peculiar shape of the two-spin energy spectrum which can generate an effective antiferromagnetic blockade. In particular, here the system would like to maximize the number of ferromagnetic domains but, in doing so, the constraint of energy conservation cannot be fulfilled. For this reason, an effective antiferromagnetic blockade is present and the number of antiferromagnetic domains is kept finite during the evolution. As a consequence, the antiferromagnetic ordering of $|\psi_{AF}\rangle$ is partially conserved, thus producing a finite $M_s(t)$ during the whole dynamics. This effective magnetic ordering is generated by the same mechanism protecting the total number of atomic pairs when ultracold bosons are abruptly driven in the strongly repulsive regime \cite{RBP1}.  
As in the previous two-spin results, Fig. \ref{fig:four} makes evident that smaller values of $\alpha$ radically affect also the many-spins dynamics. Indeed, for $\alpha=1/2,1$ $M_s(t)$ rapidly decreases and oscillates around zero thus meaning that the antiferromagnetic ordering of $|\psi_{AF}\rangle$ is lost. In these cases, the two-spin bandwidth $W(k)$ is unbounded and, as consequence, no effective antiferromagnetic blockade can be generated.
For this reason the system can reach the most favorable spin configuration where the number of ferromagnetic domains is maximized.\\ A further interesting scenario is represented by the case of infinite-range spin couplings $\alpha=0$. This regime can turn out to be of crucial relevance for experiments involving atoms in cavities where infinite range interactions are naturally present \cite{piazza}. 
Here we can derive an effective model (see appendix \ref{sec:app_fully})
\begin{equation}
H = 2\,J \left( (S)^2+(\Delta-1)(S^z)^2\right),
\label{fully}
\end{equation}
where we introduced the collective spin operator 
$S^2=(S^x)^2+(S^y)^2+(S^z)^2$, where $S^{x,y,z} = \sum_{i=1}^L S^{x,y,z}_i$. In this specific case, $M_s(t)$ 
can be written as a combination of the Clebsch-Gordan coefficients and its time evolution can be derived analytically, see the appendix \ref{sec:app_fully}. The produced evolution of the staggered magnetization for the fully connected model eq. (\ref{fully}) is shown as a green line in Fig. \ref{fig:four}.
Noticeably the dynamics turns out to be fully periodic with a period 
$T\, |J|/\hbar = \pi$ and it {\it does not} depend on the 
value of the anisotropy $\Delta$. 
Also, for $L\ge 4$ the staggered magnetization decreases to
moderately negative values and then
 it revives to the largest value $M_s(t)=1/2$.
\begin{figure}[t!]
\centering
 \includegraphics[width=.99\linewidth]{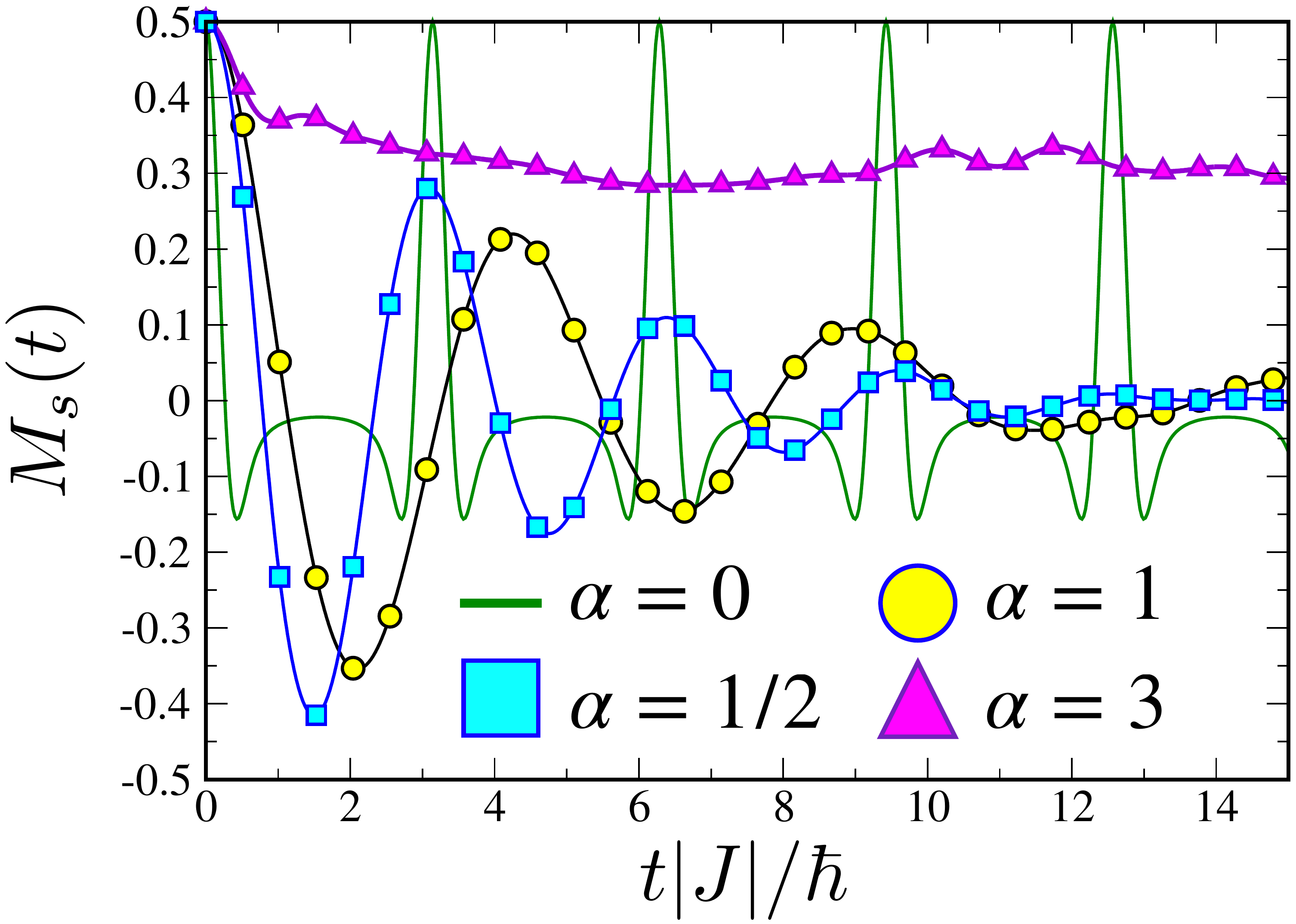}
\caption{Staggered magnetization density
$M_s(t)=\sum_i (-1)^i \left< S^z_i \right>/\, L$ 
for $\alpha=0$ (green line), $0.5$  (cyan squares), 
$\alpha=1$ (yellow circles), and $\alpha=3$ (magenta triangles) of an initial antiferromagnetic state
evolving with Hamiltonian of Eq. (\ref{hamspin}), with $\Delta/J=-5$ and $L=24$. 
For t-DMRG simulations we employ
open boundary conditions. 
The green line corresponds to the infinite range model $\alpha=0$ described in the text. For the scaling with the particle number of the $\alpha=0$ model see fig.\ref{fig:six} in the appendix.}
\label{fig:four}
\end{figure}

\section{Trapped-Ion Implementation}
\label{trappedion}
The results presented above can be accessed experimentally in
a variety of experimental platforms. 
Here we focus on a specific implementation in an array of trapped ions. Trapped ions offer pristine spin systems that can be initialized in the $\ket{\downarrow}$ state via optical pumping and interrogated with high fidelity at the end of the experiment with state-dependent fluorescence. The spin degree of freedom can be addressed with electromagnetic fields to implement rotations or entangling operations. The latter are realized by coupling the spins with normal modes of motion of the trapped-ion crystals. By off-resonantly coupling spin and motional degrees of freedom, it is possible to engineer Ising-like spin-spin interactions\cite{ions}:
\begin{equation}
H_{\gamma\gamma}=\sum_{ij} J_{ij}S^{\gamma}_i S^{{\gamma}}_j,
\label{eq_Ising_ions}
\end{equation}
where $\gamma$ can be set to $x,y,z$ depending on the experimental configuration and the spin-spin couplings can be approximately described as a power law $J_{ij}\sim J_0/|i-j|^\alpha$ with a tunable power-law exponent $\alpha\in(0,3)$ \cite{Defenu2021}. An experimental approach to simulate general XYZ Hamiltonian with trapped ions has been proposed in Ref. \cite{Bermudez2017} using quasi-periodic drives and in Ref. \cite{porras2004, deng, Davoudi2020} using orthogonal sets of normal modes of motion to create tailored two-body interactions. In the latter approach, the long-range XXZ model can be simulated in two steps: a set of radial normal modes can be used to induce an XY antiferromagnetic coupling by asymmetrically driving the spin-motion couplings \cite{ions}. This procedure creates an effective magnetic field along the $z$ direction allowing to retain only the spin preserving parts of the Ising interactions\cite{Defenu2021}. In addition, axial modes can be driven with a phase-gate type of interactions \cite{Leibfried2003} with a negative detuning to create independently tunable ferromagnetic couplings. 

An alternative approach to generate the XXZ dynamics is to approximate the evolution by means of Trotterization of the Ising Hamiltonians \cite{Lloyd1996, Lanyon2011}: applying global rotations in-between different Ising chapters will allow to effectively apply Ising terms along different directions of the Bloch sphere, obtaining an effective XXZ Hamiltonian:
\begin{equation}
e^{-i H_{\rm XXZ} t}\approx \left(e^{-i H_{xx}t/n} e^{-i H_{yy} t/n} e^{-i H_{zz}  t/n}\right)^{n}
\end{equation}
where $\tau=t/n$ is the Trotter step. The main drawbacks of this approach are the finite fidelity of the global rotations and the Trotter errors, which can be estimated as $\sum_{i>j} n \tau^2 [H_{ii},H_{jj}] + \mathcal{O}( n \tau^3)$. It is worth noting that, in order to observe the dynamical behaviour described here, it is sufficient to measure local observables, such as magnetization or two-body correlations. These observables have less stringent requirements compared to the approximation of the full unitary \cite{Heyl2019} and they are readily accessible in trapped-ion platforms that feature high fidelity individual single shot detection of all the spins along any measurement basis.

Both methods allow to tune the ratio $\Delta/J$ over the range required to observe the phenomenology discussed in this paper. Most importantly, trapped-ion systems provide the capability of tuning the long-range character of the spin-spin interactions. Even if the theoretically attainable range is $0<\alpha<3$, realistic experimental parameters are constrained between $0.5\leq\alpha\leq2$ due to practical limitations \cite{ions,Defenu2021}. However, this should be a sufficient range to observe qualitatively different dynamical behaviours as a function of system sizes and power-law exponent. Finally we stress that preparing initial product states with local spin excitations or even antiferromagnetic states is standard practice in trapped-ion systems \cite{tan2021domain, Kyprianidis2021, morong2021observation, Kokail2019}.

\section{Conclusions}
In this work we studied the effects produced by pure long-range couplings in the celebrated spin-$1/2$ XXZ Heisenberg model. We revealed how bound states in the two-spin energy spectrum can easily be generated or destroyed by simply tuning the value of the coupling decays. More precisely, we showed that pure long-range couplings occurring for $\alpha\leq1$ allow for the merging of all the energetic states into a scattering region with fully delocalized associated wave functions. On the contrary, when $\alpha>1$ defined bound states characterize the two-spin energy spectrum. As we numerically derived, these features drastically affect the two-spin dynamical properties in an energy conserving configuration. Indeed, our exact diagonalization results demonstrated that for $\alpha>1$ bound states can be dynamically populated and that they expand following the usual light-cone propagation as predicted by the Lieb-Robinson bound. On the other hand, for $\alpha\leq1$ we observe a dynamics associated with the expansion of unbounded spins with almost instantaneous propagation for any value of the $\Delta/J$. These phenomena have strong implications also for the many-spin dynamics. Indeed, we showed that on one hand for $\alpha<1$ antiferromagnetic orders are dynamically destroyed for ferromagnetic couplings. On the other hand when $\alpha>1$ the combination of antiferromagnetic blockade and energy conservation can support the presence of a novel long-lived antiferromagnetic state in presence of a strong ferromagnetic coupling. Moreover, in the extreme regime of infinite-range coupling $\alpha=0$, the anisotropy along the $z$-direction does not play any role, and the (staggered) magnetization leads to a periodic oscillatory behavior. Finally, despite the fact that our results for $\alpha=0,\, 3$ might be tested in experimental platforms involving atoms in cavities and polar molecules or Rydberg atoms respectively, we devised an experimental scheme based on a trapped-ion quantum simulator. The latter could pave the way toward the experimental realization of the long-range XXZ Hamiltonian and the experimental investigation of all the aforementioned effects. Our experimental proposal makes use of the tunability of trapped-ion quantum simulators and it represents a relevant step forward toward a systematic and versatile quantum simulation of long-range interacting quantum systems. It is also worth to underline that, as a natural extension of our results, it would be interesting to study similar effects in spin-$1$ systems where topological phases exist both at short- \cite{haldane} and long-range level \cite{gong}. Moreover, a further interesting direction would be the investigation of the two-spin properties of constrained long-range spin models \cite{bluvstein} which describe lattice gauge theories \cite{borla,celi,kerbic}

\section*{Acknowledgements}
We thank Domenico Giuliano, Samihr Hermes, Chiara Menotti and Simone Paganelli for useful discussions and the High Performance Computing Center (NPAD) at UFRN for providing computational resources. L. L. acknowledges  financial support by the PRIN project number 20177SL7HC, financed by the Italian Ministry of education and research.
T. M. acknowledges CNPq for support through Bolsa de 
produtividade em Pesquisa n.311079/2015-6 and the Serrapilheira Institute (Grant No. 
Serra-1812-27802), CAPES-NUFFIC Project No. 88887.156521/2017-00. G. P. acknowledges support by the DOE Office of Science, Office of Nuclear Physics, under Award no. DE-SC0021143, the Office of Naval Research (N00014-20-1-2695), the Army Research Lab (W911QX20P0063) and the Army Research office (W911NF21P0003) 
M. L. and L. B. acknowledge support from ERC AdG NOQIA, State Research Agency AEI (“Severo Ochoa” Center of Excellence CEX2019-000910-S, Plan National FIDEUA PID2019-106901GB-I00/10.13039 / 501100011033, FPI, QUANTERA MAQS PCI2019-111828-2 / 10.13039/501100011033), Fundació Privada Cellex, Fundació Mir-Puig, Generalitat de Catalunya (AGAUR Grant No. 2017 SGR 1341, RETOS-QUSPIN, CERCA program, QuantumCAT \ U16-011424, co-funded by ERDF Operational Program of Catalonia 2014-2020), EU Horizon 2020 FET-OPEN OPTOLogic (Grant No 899794), and the National Science Centre, Poland (Symfonia Grant No. 2016/20/W/ST4/00314), Marie Sk\l odowska-Curie grant STREDCH No 101029393, “La Caixa” Junior Leaders fellowships (ID100010434),  and EU Horizon 2020 under Marie Sk\l odowska-Curie grant agreement No 847648 (LCF/BQ/PI19/11690013, LCF/BQ/PI20/11760031,  LCF/BQ/PR20/11770012). T.M. thanks for the hospitality the Physics Department of the University
of L'Aquila, where part of the work has been done.

\newpage

\onecolumngrid

\appendix

\section{Coupling coefficients of the effective Hamiltonian Eq.(\ref{Heff_1})}
\label{app2}
For initial states where both spins sitting next to each others and isolated spins are present, a specific quench protocol consisting in making this states evolve with Eq.\ref{hamspin} for $|\Delta|\gg J$ allows to derive the effective Hamiltonian:
\begin{eqnarray}
H^{(\mathrm{eff})}&&=\sum_{i<j} (\tilde{t}^{(D)}_{ij} \, \tilde{D}_i^\dagger \tilde{D}_j + \tilde{w}^{(sD)}_{ij} \, \tilde{S}_i^{+} \tilde{D}_j^{\dagger} \, \tilde{D}_i \tilde{S}^-_j+\nonumber\\
&&+\frac{J\tilde{S}^+_i\tilde{S}^-_j}{2|i-j|^{\alpha}}+\mathrm{H. c.} )
\label{Heff}
\end{eqnarray}
where $\tilde{t}^{(D)}_{ij}$ describes the hopping amplitude of two spins forming a bound state, $J$ is the usual single spin spreading velocity and $\tilde{w}^{(sD)}_{ij}$ refers to the possible swapping processes occurring between bounded and unbounded spins. By means of perturbation theory up to the second order we can derive explicitly the aforementioned parameters. In particular our calculation shows that
\begin{align}
\tilde{t}^{(D)}_{ij} = \Big(\frac{J}{2^{\alpha+1}} + w_1 \, \frac{J}{4\Delta}\Big) \, \delta_{d_{ij},1} + w_2 \, \frac{J}{2\Delta} \, \delta_{d_{ij},2} + \quad \quad \quad \quad \quad \quad \quad \quad \quad \quad \quad \nonumber \\
{} \label{tdob} \\
+ \frac{J}{4\Delta} \, \Bigg\{ \frac{1}{d_{ij}^{2 \alpha}} \, \Bigg[ \frac{(d_{ij}+1)^{\alpha}}{(d_{ij}+1)^{\alpha} -1}  + \frac{(d_{ij} -1)^{\alpha}}{(d_{ij} -1)^{\alpha} -1} \, (1- \delta_{d_{ij},2}) \Bigg]  +  \frac{2}{(d_{ij}^2-1)^{\alpha}} \, \frac{d_{ij}^{\alpha}}{d_{ij}^{\alpha} -1 } \, (1- \delta_{d_{ij},1}) \Bigg\} \, , \nonumber
\end{align}
with
\begin{equation}
 w_1 = 
  \sum_{n = 3}^{\infty} \frac{1}{[n (n - 2)]^{\beta}} \Bigg(\frac{(n -1)^{\alpha}}{(n -1)^{\alpha} -1} \Bigg)  +  \sum_{n = 1}^{\infty} \frac{1}{[n (n + 2)]^{\alpha}} \Bigg(\frac{(n  + 1)^{\alpha}}{(n + 1)^{\alpha} -1} \Bigg)\, ,
 \quad 
 w_2 = 
  \sum_{n = 1}^{2} \frac{1}{[n (n - 2)]^{\alpha}} \Bigg(\frac{(n -1)^{\alpha}}{(n - 1)^{\alpha} -1} \Bigg)\, ,
\label{tdob12}
\end{equation}
and 
\begin{align}
\tilde{w}^{(sD)}_{ij} = \frac{J}{2(d_{ij} -1)^{\alpha}} \,  \delta_{d_{i,j},3} + (1-  \delta_{d_{i,j},1} -  \delta_{d_{i,j},2}-  \delta_{d_{i,j},3}) \,  \frac{J}{4\Delta} \, \Bigg\{ \sum_{m = d_{ij} + 2}^{\infty} \frac{1}{[m \, (m - d_{ij} +1)]^{\alpha}} \frac{1}{|1 + \frac{1}{d_{ij}^{\alpha}} - \frac{1}{(m+1)^{\alpha}} - \frac{1}{(m - d_{ij})^{\alpha}}|}  \, +  \nonumber \\
{} \label{scambiodob} \\
+  \sum_{m = 1}^{d_{ij} - 2} \frac{1}{[m \, (d_{ij} -m-1)]^{\beta}} \frac{1}{|1 + \frac{1}{d_{ij}^{\alpha}} - \frac{1}{(m+1)^{\alpha}} - \frac{1}{(d_{ij} - m)^{\alpha}}   | }   +  \sum_{m = 3}^{\infty} \frac{1}{m \, (d_{ij} + m-1)]^{\alpha}} \frac{1}{|1 + \frac{1}{d_{ij}^{\alpha}} - \frac{1}{(m - 1)^{\alpha}} - \frac{1}{(d_{ij} + m)^{\alpha}}|}  \Bigg\} \, . \nonumber
\end{align} 

\section{Dynamics of the antiferromagnetic state for infinite-range couplings ($\alpha \to 0$).}
\label{sec:app_fully}

\begin{figure*}[t!]
\includegraphics[width=0.6\linewidth]{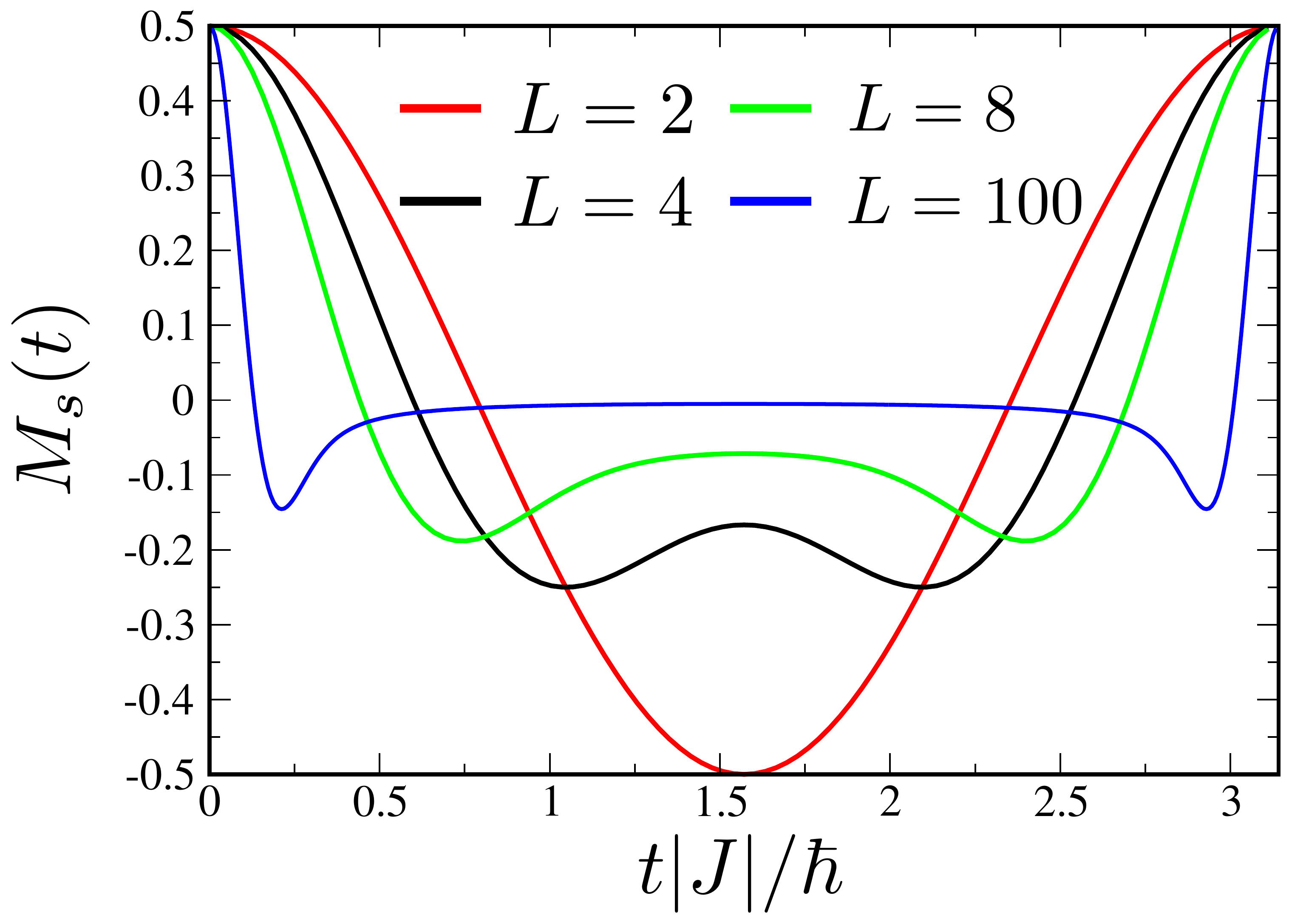}
\caption{Dynamics of the staggered magnetization $M_s(t)=\sum_i (-1)^i \left< S^z_i \right>/\, L$ 
for different lengths of the chain for infinite range couplings of Eq.(\ref{eq:fully_coll}). $L=2$ (red line), $L=4$  (black), $L=8$ (green), and $L=100$ (black).
The dynamics for the model is periodic, with period $T J/\hbar=\pi$.
} 
\label{fig:six}
\end{figure*}

In this appendix we compute the staggered 
magnetization for the initial antiferromagnetic state
\beq
|\psi_{AF}\rangle=|\downarrow_1\uparrow_2...\downarrow_{L/2}\uparrow_{L/2+1}...\downarrow_{L-1}\uparrow_L\rangle \label{eq:neel}
\eeq
evolving with the fully connected Hamiltonian
\beq
\label{fully_conn}
H = 2\,J \sum_{i<j} S^x_i S^x_j + S^y_i S^y_j + \Delta S^z_i S^z_j.
\eeq
Eq.(\ref{fully_conn}) corresponds to the special case $\alpha=0$ in Eq.(\ref{hamspin}) and it can be rewritten (removing an irrelevant constant) as
\beq
\label{eq:fully_coll}
H =2\, J \left( S^2+(\Delta-1)S_z^2\right),
\eeq
where we introduced the total spin operator $S=(S^x)^2+(S^y)^2+(S^z)^2$, and $S^{x,y,z}= \sum_{i=1}^L S^{x,y,z}_i$ are 
collective spin operators.
To evolve the antiferromagnetic state of Eq.(\ref{eq:neel}) we need to express it in terms of the eigenstates of the collective spin operators. 
We observe that the state Eq.(\ref{eq:neel}) can be written as the direct product of two polarized states with spin projection $(L/2,-L/2)$ for the two sublattices (the first with odd lattice indexes, the second with even indexes) in which the chain can be divided.
The state of Eq.(\ref{eq:neel}) then evolves as
\beq
\label{eq:dyn_AF}
|\psi_{AF}(t)\rangle=
\sum_{j=0}^{L/2}
\sqrt{\frac{\frac{L}{2}!(2j+1)}{(\frac{L}{2}+j+1)!(\frac{L}{2}-j)!}} e^{-i 2\,J\, t\, j(j+1)/\hbar}
|j,0\rangle,
\eeq
where the states $|j,0\rangle$ are the eigenstates of 
the collective spin operators $(S^2,S^z)$.
The term $(S^z)^2$ of Eq.(\ref{eq:fully_coll}) does not affect the dynamics, since the third componente of the total spin acts as $S^z |\psi_{AF}\rangle=0$. Expressing the collective states $|j,0\rangle$ in terms of the basis
\beq
\left| \frac{L}{2},m;\frac{L}{2},-m\right>,
\eeq
where the first (second) pair of quantum number refers to the first (second) sublattice, we can rewrite the time-evolved state of Eq.(\ref{eq:dyn_AF}) in a form which is better suited to compute the time-dependent  staggered magnetization $M_s(t)=\sum_i (-1)^i \left< S^z_i \right>/\, L$:
\beq
|\psi_{AF}(t)\rangle=
\sum_{m=-L/4}^{L/4} c_m^{2L}(t) 
\left| \frac{L}{2},m;\frac{L}{2},-m\right>,
\eeq
where we introduced the time-dependent coefficients
\beq
c_m^{2L}(t) = 
\sum_{j=0}^{L/2}
\sqrt{\frac{\frac{L}{2}!(2j+1)}{(\frac{L}{2}+j+1)!(\frac{L}{2}-j)!}} e^{-i 2\,J\, t\, j(j+1)/\hbar}
\left< \frac{L}{2},m;\frac{L}{2},-m |j,0\right>.
\eeq

The staggered magnetization can then be written as 
\beq
M_s(t) = \sum_{m=-L/4}^{L/4} 2m\, |c_m^{2L}(t)|^2.
\eeq
This result depends on the number of spins $L$, but it does not depend on the anisotropy $\Delta$. In fig.\ref{fig:six} we plot the time evolution of the staggered magnetization for different particle numbers up to $t\,|J|/\hbar = \pi$. We notice that the dynamics is periodic, since the frequencies involved are integer numbers $j(j+1)$ with $j=0,1,\dots L/2$. 
Finally we observe that only for $L=2$ does the system oscillates between the two antiferromagnetic states
$|\psi_{AF}\rangle=|\downarrow_1\uparrow_2...\downarrow_{L/2}\uparrow_{L/2+1}...\downarrow_{L-1}\uparrow_L\rangle$ and 
$|\psi_{AF}'\rangle=|\uparrow_1\downarrow_2...\uparrow_{L/2}\downarrow_{L/2+1}...\uparrow_{L-1}\downarrow_L\rangle$. Also, for $L\ge 4$ the staggered magnetization decreases to moderately negative values and then it revives to the largest value $M_s=1/2$.

\twocolumngrid


\begin{thebibliography}{99}
\bibliographystyle{unsrt}
\bibitem{silva} 
A. Polkovnikov, K. Sengupta, A. Silva, and M. Vengalattore,
{\it Colloquium: Nonequilibrium dynamics of closed interacting quantum systems},
\href{https://journals.aps.org/rmp/abstract/10.1103/RevModPhys.83.863}{Rev. Mod. Phys. {\bf 83}, 863 (2011)}.

\bibitem{rigol}
M. Rigol, V. Dunjko, and M. Olshanii, 
{\it Thermalization and its mechanism for generic isolated quantum systems},
\href{https://www.nature.com/articles/nature06838}{Nature {\bf 452}, 854 (2008)}.

\bibitem{gogolin}
C. Gogolin and J. Eisert,
{\it Equilibration, thermalisation, and the emergence of statistical mechanics in closed quantum systems},
\href{https://iopscience.iop.org/article/10.1088/0034-4885/79/5/056001}{Rep. Prog. Phys. {\bf 79}, 056001 (2016)}.

\bibitem{dalessio}
L. D' Alessio, Y. Kafri, A. Polkovnikov, and M. Rigol
{\it From Quantum Chaos and Eigenstate Thermalization to Statistical Mechanics and Thermodynamics},
\href{https://www.tandfonline.com/doi/full/10.1080/00018732.2016.1198134}{Adv. Phys. {\bf 65}, 239 (2016)}.

\bibitem{essler2}
F. H. L. Essler and M. Fagotti,
{\it Quench dynamics and relaxation in isolated integrable quantum spin chains},
\href{https://iopscience.iop.org/article/10.1088/1742-5468/2016/06/064002}{J. Stat. Mech. 064002 (2016)}.

\bibitem{goldman}
N. Goldman and J. Dalibard,
{\it Periodically Driven Quantum Systems: Effective Hamiltonians and Engineered Gauge Fields},
\href{https://journals.aps.org/prx/abstract/10.1103/PhysRevX.4.031027}{Phys. Rev. X {\bf 4}, 031027 (2014)}.

\bibitem{bukov}
M. Bukov, L. D' Alessio, and A. Polkovnikov,
{\it Universal high-frequency behavior of periodically driven systems: from dynamical stabilization to Floquet engineering},
\href{https://www.tandfonline.com/doi/full/10.1080/00018732.2015.1055918}{Adv. Phys. {\bf 64}, 139–226 (2015)}.

\bibitem{eckardt} 
A. Eckardt, 
{\it Colloquium: Atomic quantum gases in periodically driven optical lattices},
\href{https://journals.aps.org/rmp/abstract/10.1103/RevModPhys.89.011004}{Rev. Mod. Phys. {\bf 89}, 011004 (2017)}.

\bibitem{barbiero} 
L. Barbiero, C. Schweizer, M. Aidelsburger, E. Demler, N.Goldman, and F. Grusdt, 
{\it Coupling ultracold matter to dynamical gauge fields in optical lattices: From flux attachment to $Z_2$ lattice gauge theories}, \href{https://advances.sciencemag.org/content/5/10/eaav7444}{Sc. Adv. {\bf5}, 10 (2019)}.

\bibitem{tilman} 
F. G\"org, K. Sandholzer, J. Minguzzi, R. Desbuquois, M. Messer, and T. Esslinger, 
{\it Realization of density-dependent Peierls phases to engineer quantized gauge fields coupled to ultracold matter},
\href{https://www.nature.com/articles/s41567-019-0615-4}{Nat. Phys. {\bf 15}, 1 (2019)}.

\bibitem{monika} 
C. Schweizer, F. Grusdt, M. Berngruber, L. Barbiero, E. Demler, N. Goldman, I. Bloch, and M. Aidelsburger, 
{\it Floquet approach to $Z_2$ lattice gauge theories with ultracold atoms in optical lattices},
\href{https://www.nature.com/articles/s41567-019-0649-7}{Nature Physics 15, 1168-1173 (2019)}.

\bibitem{greiner} 
M. E. Tai, A. Lukin, M. Rispoli, R. Schittko, T. Menke, D. Borgnia, P. M. Preiss, F. Grusdt, A. M. Kaufman, and M. Greiner, 
{\it Microscopy of the interacting Harper–Hofstadter model in the two-body limit},
\href{https://www.nature.com/articles/nature22811}{Nature {\bf 546}, 519-523 (2017)}.

\bibitem{das} 
A. Das and B. K. Chakrabarti,
{\it Colloquium: Quantum annealing and analog quantum computation},
\href{https://journals.aps.org/rmp/abstract/10.1103/RevModPhys.80.1061}{Rev. Mod. Phys. {\bf 80}, 3 1061 (2008)}.

\bibitem{santoro}
G. E. Santoro, R. Martonak, E. Tosatti, and R. Car,
{\it Theory of Quantum Annealing of an Ising Spin Glass},
\href{https://science.sciencemag.org/content/295/5564/2427}{Science {\bf 295}, 2427 (2002).}

\bibitem{albash}
T. Albash and D. A. Lidar,
{\it Adiabatic Quantum Computation},
\href{https://journals.aps.org/rmp/abstract/10.1103/RevModPhys.90.015002}{Rev. Mod. Phys. {\bf 90}, 015002 (2018)}.

\bibitem{lukin} 
A. Keesling, A. Omran, H. Levine, H. Bernien, H. Pichler, S. Choi, R. Samajdar, S. Schwartz, P. Silvi, S. Sachdev, P. Zoller, M. Endres, M. Greiner, V. Vuletic, and M. D. Lukin,
{\it Quantum Kibble–Zurek mechanism and critical dynamics on a programmable Rydberg simulator},
\href{https://www.nature.com/articles/s41586-019-1070-1}{Nature {\bf 568}, 207 (2019)}.

\bibitem{calabrese} 
P. Calabrese and J. Cardy,
{\it Evolution of entanglement entropy in one-dimensional systems},
\href{https://iopscience.iop.org/article/10.1088/1742-5468/2005/04/P04010}{J. Stat. Mech. P04010 (2004)}.


\bibitem{bloch}
M. Schreiber, S. S. Hodgman, P. Bordia, H. P. Lüschen, M. H. Fischer, R. Vosk, E. Altman, U. Schneider, and I. Bloch,
{\it Observation of many-body localization of interacting fermions in a quasi-random optical lattice},
\href{https://science.sciencemag.org/content/349/6250/842}{Science {\bf 349}, 842 (2015)}.

\bibitem{nagerl} 
F. Meinert, M. J. Mark, E. Kirilov,
K. Lauber, P. Weinmann, M. Gröbner, A. J. Daley, and H.-C. Nägerl, 
{\it Observation of many-body dynamics in long-range tunneling after a quantum quench},
\href{https://science.sciencemag.org/content/344/6189/1259.abstract}{Science {\bf 344}, 1259-1262 (2014)}.

\bibitem{pichler} 
T. Pichler, M. Dalmonte, E. Rico, P. Zoller, and S. Montangero,
{\it Real-Time Dynamics in U(1) Lattice Gauge Theories with Tensor Networks},
\href{https://journals.aps.org/prx/abstract/10.1103/PhysRevX.6.011023}{Phys. Rev. X {\bf 6}, 011023 (2016)}.

\bibitem{barbiero2018}
L. Barbiero, L. Santos, and N. Goldman
{\it Quenched dynamics and spin-charge separation in an interacting topological lattice},
\href{https://doi.org/10.1103/PhysRevB.97.201115}{Phys. Rev. B {\bf 97}, 201115 (2018)}.

\bibitem{monika2} 
T. Kohlert, S. Scherg, P. Sala, F. Pollmann, B. H. Madhusudhana, I. Bloch, and M. Aidelsburger,
{\it Experimental realization of fragmented models in tilted Fermi-Hubbard chains},
\href{https://arxiv.org/abs/2106.15586}{arXiv:2106.15586}.


\bibitem{RBP1} 
K. Winkler, G. Thalhammer, F. Lang, R. Grimm, J. Hecker-Denschlag, A. J. Daley, A. Kantian, H. P. Buchler, and P. Zoller, 
{\it Repulsively bound atom pairs in an optical lattice},
\href{https://www.nature.com/articles/nature04918}{Nature {\bf 441}, 853 (2006)}.

\bibitem{RBP2} 
N. Strohmaier, D. Greif, R. J\"ordens, L. Tarruell, H. Moritz, T. Esslinger, R. Sensarma, D. Pekker, E. Altman, and E. Demler,
{\it Observation of Elastic Doublon Decay in the Fermi-Hubbard Model},
\href{https://journals.aps.org/prl/abstract/10.1103/PhysRevLett.104.080401}{Phys. Rev. Lett {\bf 104}, 080401 (2010)}.

\bibitem{io} 
L. Barbiero, C. Menotti, A. Recati, and L. Santos,
{\it Out-of-equilibrium states and quasi-many-body localization in polar lattice gases}, 
\href{https://journals.aps.org/prb/abstract/10.1103/PhysRevB.92.180406}{Phys. Rev. B {\bf 92}, 180406 (2015)}. 

\bibitem{li2020}
W. Li, A. Dhar, X. Deng, K. Kasamatsu, L. Barbiero, and L. Santos,
{\it Disorderless Quasi-localization of Polar Gases in One-Dimensional Lattices},
\href{https://journals.aps.org/prl/abstract/10.1103/PhysRevLett.124.010404}{Phys. Rev. Lett. {\bf 124}, 010404 (2020)}.

\bibitem{muth} 
D. Muth, D. Petrosyan, and M. Fleischhauer,
{\it Dynamics and evaporation of defects in Mott-insulating clusters of boson pairs},
\href{https://journals.aps.org/pra/abstract/10.1103/PhysRevA.85.013615}{Phys. Rev. A {\bf 85}, 013615 (2012)}.

\bibitem{nagerl2} 
M. J. Mark, E. Haller, K. Lauber, J. G. Danzl, A. Janisch, H. P. Buchler, A. J. Daley, and H.-C. Nagerl, 
{\it Preparation and Spectroscopy of a Metastable Mott-Insulator State with Attractive Interactions},
\href{https://journals.aps.org/prl/abstract/10.1103/PhysRevLett.108.215302}{Phys. Rev. Lett. {\bf 108}, 215302 (2012)}.

\bibitem{fukuhara} 
T. Fukuhara, P. Schauss, M. Endres, S. Hild, M. Cheneau, I. Bloch, and C. Gross, 
{\it Microscopic observation of magnon bound states and their dynamics},
\href{https://www.nature.com/articles/nature12541}{Nature {\bf 502}, 76 (2013)}.

\bibitem{ketterle}
P. N. Jepsen, W. W. Ho, J. Amato-Grill, I. Dimitrova, E. Demler, and W. Ketterle
{\it Transverse spin dynamics in the anisotropic Heisenberg model realized with ultracold atoms},
\href{https://arxiv.org/abs/2103.07866}{arXiv:2103.07866}.

\bibitem{giuliano2013}
D. Giuliano, D. Rossini, P. Sodano, and A. Trombettoni,
{\it XXZ spin-$\frac{1}{2}$ representation of a finite-U Bose-Hubbard chain at half-integer filling},
\href{https://journals.aps.org/prb/abstract/10.1103/PhysRevB.87.035104}{Phys. Rev. B {\bf 87}, 035104 (2013)}.

\bibitem{book_magnetism}
{\it Quantum Magnetism},
Editors: U. Schollw\"ock, J. Richter, D. J. J. Farnell, R. F. Bishop,
Springer (2004).

\bibitem{essler} 
M. Ganahl, E. Rabel, F. H. L. Essler, and H. G. Evertz, 
{\it Observation of Complex Bound States in the Spin-$1/2$ Heisenberg $XXZ$ Chain Using Local Quantum Quenches},
\href{https://journals.aps.org/prl/abstract/10.1103/PhysRevLett.108.077206}{Phys. Rev. Lett. {\bf 108}, 077206 (2012)}.

\bibitem{io3}
L. Barbiero and L. Dell'Anna, 
{\it Spreading of correlations in a quenched repulsive and attractive one-dimensional integrable system},
\href{https://journals.aps.org/prb/abstract/10.1103/PhysRevB.96.064303}{Phys. Rev. B {\bf 96}, 064303 (2017)}.

\bibitem{io2} 
L. Barbiero, L. Chomaz, S. Nascimbene, and N. Goldman, 
{\it Bose-Hubbard physics in synthetic dimensions from interaction Trotterization},
\href{https://journals.aps.org/prresearch/abstract/10.1103/PhysRevResearch.2.043340}{Phys. Rev. Res. {\bf 2}, 043340 (2020)}.

\bibitem{altman} 
P. Barmettler, M. Punk, V. Gritsev, E. Demler and E. Altman,
{\it Relaxation of Antiferromagnetic Order in Spin-1/2 Chains Following a Quantum Quench},
\href{https://journals.aps.org/prl/abstract/10.1103/PhysRevLett.102.130603}{Phys. Rev. Lett. {\bf 102}, 130603 (2009)}.

\bibitem{dauxois}
T. Dauxois, S. Ruffo, E. Arimondo, and M. Wilkens,
{\it Dynamics and thermodynamics of systems with Long Range Interactions},
Lecture Notes in Physics (Springer- Verlag, Berlin Heidelberg, 2002).

\bibitem{kestner1}
M. Kastner,
{\it Nonequivalence of ensembles for long-range quantum spin systems in optical lattices},
\href{https://journals.aps.org/prl/abstract/10.1103/PhysRevLett.104.240403}{Phys. Rev. Lett. {\bf 104}, 240403 (2010)}.

\bibitem{hermes2020}
S. Hermes, T. J. G. Apollaro, S. Paganelli, and T. Macr\`i,
{\it Dimensionality-enhanced quantum state transfer in long-range interacting spin systems},
\href{https://journals.aps.org/pra/abstract/10.1103/PhysRevA.101.053607}{Phys. Rev. A {\bf 101}, 053607 (2020)}.


\bibitem{CDP2} 
M. B. Hastings, 
{\it Locality in Quantum and Markov Dynamics on Lattices and Networks},
\href{https://journals.aps.org/prl/abstract/10.1103/PhysRevLett.93.140402}{Phys. Rev. Lett. {\bf 93}, 140402 (2004)}.

\bibitem{CDP3}
B. Nachtergaele and R. Sims, 
{\it Lieb-Robinson Bounds and the Exponential Clustering Theorem},
\href{https://link.springer.com/article/10.1007/s00220-006-1556-1}
{Comm. Math. Phys. {\bf 265}, 119 (2006)}.

\bibitem{LRB} 
M. C. Tran, C. F. Chen, A. Ehrenberg, A. Y. Guo, A. Deshpande, Y. Hong, Z.-X. Gong, A. V. Gorshkov, and A. Lucas, 
{\it Hierarchy of Linear Light Cones with Long-Range Interactions},
\href{https://journals.aps.org/prx/abstract/10.1103/PhysRevX.10.031009}{Phys. Rev. X {\bf 10}, 031009 (2020)}.
\bibitem{LRB1} 
P. Calabrese and J. Cardy, 
{\it Time Dependence of Correlation Functions Following a Quantum Quench},
\href{https://journals.aps.org/prl/abstract/10.1103/PhysRevLett.96.136801}{Phys. Rev. Lett. {\bf 96}, 136801 (2006)}.

\bibitem{LRB2}
E. H. Lieb and D. W. Robinson, 
{\it The finite group velocity of quantum spin systems},
\href{https://link.springer.com/article/10.1007/BF01645779}{Comm. Math. Phys. {\bf 28}, 251 (1972)}.

\bibitem{AL1} 
A. Osterloh, L. Amico, G. Falci, and R. Fazio, {\it Scaling of entanglement close to
a quantum phase transition}, 
\href{https://www.nature.com/articles/416608a}{Nature {\bf 416}, 6881, 608--610 (2002)}.

\bibitem{AL2} 
J. Eisert and M. Cramer, and M. B. Plenio,
{\it Area laws for the entanglement entropy, Reviews of Modern Physics},
\href{https://journals.aps.org/rmp/abstract/10.1103/RevModPhys.82.277}{Rev. Mod. Phys. {\bf 82}, 277 (2010)}.

\bibitem{hauke2013}
P. Hauke and L. Tagliacozzo,
{\it Spread of Correlations in Long-Range Interacting Quantum Systems},
\href{https://journals.aps.org/prl/abstract/10.1103/PhysRevLett.111.207202}{Phys. Rev. Lett. {\bf 111}, 207202 (2013)}.

\bibitem{maghrebi}
M. F. Maghrebi, Z.-X. Gong, and A. V. Gorshkov,
{\it Continuous Symmetry Breaking in 1D Long-Range Interacting Quantum Systems},
\href{https://journals.aps.org/prl/abstract/10.1103/PhysRevLett.119.023001}{Phys. Rev. Lett. {\bf 119}, 023001 (2017)}.

\bibitem{kastner2014}
J. Eisert, M. van den Worm, S. R. Manmana, and M. Kastner,
{\it Breakdown of Quasilocality in Long-Range Quantum Lattice Models},
\href{https://journals.aps.org/prl/abstract/10.1103/PhysRevLett.111.260401}{Phys. Rev. Lett. {\bf 111}, 260401 (2013)}.

\bibitem{gorshkov2016}
M. F. Maghrebi, Z.-X. Gong, M. Foss-Feig, and A. V. Gorshkov,
{\it Causality and quantum criticality in long-range lattice models},
\href{https://journals.aps.org/prb/abstract/10.1103/PhysRevB.93.125128}{Phys. Rev. B {\bf 93}, 125128 (2016)}.


\bibitem{daley2016}
A. S. Buyskikh, M. Fagotti, J. Schachenmayer, F. Essler, and A. J. Daley,
{\it Entanglement growth and correlation spreading with variable-range interactions in spin and fermionic tunneling models},
\href{https://journals.aps.org/pra/abstract/10.1103/PhysRevA.93.053620}{Phys. Rev. A {\bf 93}, 053620 (2016)}.

\bibitem{lepori2017}
L. Lepori, A. Trombettoni, and D. Vodola, 
{\it Singular dynamics and emergence of nonlocality in long-range quantum models}, 
\href{https://iopscience.iop.org/article/10.1088/1742-5468/aa569d}{Journ. Stat. Mech. 033102 (2017)}.


\bibitem{vodola2014}
D. Vodola, L. Lepori, E. Ercolessi, A. V. Gorshkov, and G. Pupillo, 
{\it Kitaev chains with long-range pairing},
\href{https://journals.aps.org/prl/abstract/10.1103/PhysRevLett.113.156402}{Phys. Rev. Lett. {\bf 113}, 156402 (2014)}.


\bibitem{vodola2016}
D. Vodola, L. Lepori, E. Ercolessi, and G. Pupillo,
{\it Long-range Ising and Kitaev Models: Phases, Correlations and Edge Modes},
\href{https://iopscience.iop.org/article/10.1088/1367-2630/18/1/015001}{New J. Phys. {\bf 18}, 015001 (2016)}.

\bibitem{pezze2018}
M. Gabbrielli, L. Lepori, and L. Pezz\`e,
{\it Multipartite-Entanglement Tomography of a Quantum Simulator},
\href{https://iopscience.iop.org/article/10.1088/1367-2630/aafb8c}{New J. Phys. {\bf 21}  033039 (2019)}.

\bibitem{lepori2018}
 L. Lepori, D. Giuliano, and S. Paganelli, 
 {\it Edge insulating topological phases in a two-dimensional long-range superconductor},
\href{https://doi.org/10.1103/PhysRevB.97.041109}{Phys. Rev. B {\bf 97}, 041109(R) (2018)}.

\bibitem{koffel2012}
T. Koffel, M. Lewenstein, and L. Tagliacozzo, 
{\it Entanglement Entropy for the Long-Range Ising Chain in a Transverse Field},
\href{https://doi.org/10.1103/PhysRevLett.109.267203}{Phys. Rev. Lett. {\bf 109}, 267203 (2012)}.

\bibitem{lepori2016}
L. Lepori, D. Vodola, G. Pupillo, G. Gori, and A. Trombettoni, 
{\it Effective theories and breakdown of conformal symmetry in a long-range quantum chain},
\href{https://doi.org/10.1016/j.aop.2016.07.026}{Ann. Phys. {\bf 374}, 35-66 (2016)}.

\bibitem{Defenu2021}
N. Defenu, T. Donner, T. Macrì, G. Pagano, S. Ruffo, and A. Trombettoni, 
{\it Long-range interacting quantum systems}, 
\href{https://arxiv.org/abs/2109.01063}{arXiv:2109.01063 (2021)}.

\bibitem{gori2013}
G. Gori, T. Macr\`i, and A. Trombettoni,
{\it Modulational instabilities in lattices with power-law hoppings and interactions},
\href{https://journals.aps.org/pre/abstract/10.1103/PhysRevE.87.032905}{Phys. Rev. E {\bf 87}, 032905 (2013)}.

 \bibitem{leporiLR}
 L. Lepori and L. Dell'Anna, 
 {\it Long-range topological insulators and weakened bulk-boundary correspondence}, 
\href{https://iopscience.iop.org/article/10.1088/1367-2630/aa84d0}{New J. Phys. {\bf 19}, 103030 (2017)}.

\bibitem{pfau}
T. Lahaye, C. Menotti, L. Santos, M. Lewenstein, and T. Pfau, 
{\it The physics of dipolar bosonic quantum gases},
\href{https://iopscience.iop.org/article/10.1088/0034-4885/72/12/126401}{Rep. Prog. Phys. {\bf 72}, 126401 (2009)}.

\bibitem{Bohn2017}
J. L. Bohn, A. M. Rey, and J. Ye, 
{\it Cold molecules: Progress in quantum engineering of chemistry and quantum matter},
\href{https://www.science.org/doi/10.1126/science.aam6299}{Science {\bf 357}, 1002 (2017)}.

\bibitem{risch} 
H. Ritsch, P. Domokos, F. Brennecke, and T. Esslinger,
{\it Cold atoms in cavity-generated dynamical optical potentials},
\href{https://journals.aps.org/rmp/abstract/10.1103/RevModPhys.85.553}{Rev. Mod. Phys. {\bf 85}, 553-601 (2013)}.

\bibitem{schauss2015} 
P. Schau\ss, J. Zeiher, T. Fukuhara, S. Hild, M. Chenau, T. Macr\`i, T. Pohl, I. Bloch, C. Gross, {\it Crystallization in Ising quantum magnets},
\href{https://science.sciencemag.org/content/347/6229/1455}{Science {\bf 347}, 6229 1455-1458 (2015)}.

\bibitem{labuhn2016}
H. Labuhn, D. Barredo, S. Ravets, S. de Léseléuc, T. Macr\`i, T. Lahaye, A. Browaeys,
{\it A highly-tunable quantum simulator of spin systems using two-dimensional arrays of single Rydberg atoms},
\href{https://www.nature.com/articles/nature18274}{Nature {\bf 534}, 667 (2016)}.

\bibitem{browaeys2020} 
A. Browaeys and T. Lahaye, 
{\it Many-body physics with individually controlled Rydberg atoms},
\href{https://www.nature.com/articles/s41567-019-0733-z}{Nat. Phys. {\bf16}, 2 (2020)}.

\bibitem{morgado2021} 
M. Morgado and S. Whitlock, 
{\it Quantum simulation and computing with Rydberg-interacting qubits},
\href{https://avs.scitation.org/doi/10.1116/5.0036562}{Quant. Sci. {\bf 3}, 023501 (2021)}.

\bibitem{ions} 
C. Monroe, W. C. Campbell, L. M. Duan, Z. X. Gong, A. V. Gorshkov, P. Hess, R. Islam, K. Kim, N. Linke, G. Pagano, P. Richerme, C. Senko, and N. Y. Yao,  
{\it Programmable quantum simulations of spin systems with trapped ions},
\href{https://journals.aps.org/rmp/abstract/10.1103/RevModPhys.93.025001}{Rev. Mod. Phys. {\bf 93}, 25001 (2021)}.

\bibitem{Britton2012}
J. W. Britton, B. C. Sawyer, A. C. Keith, J. C.-C. Wang, J. K. Freericks, H. Uys, M. J. Biercuk, and J. J. Bollinger, 
{\it Engineered two-dimensional Ising interactions in a trapped-ion quantum simulator with hundreds of spins}, 
\href{https://www.nature.com/articles/nature10981}{Nature {\bf 484}, 489 (2012)}.

\bibitem{LGT1}
E. A. Martinez, C. A. Muschik, P. Schindler, D. Nigg, A. Erhard, M. Heyl, P. Hauke, M. Dalmonte, T. Monz, P. Zoller, and R. Blatt,
{\it Real-time dynamics of lattice gauge theories with a few-qubit quantum computer},
\href{https://www.nature.com/articles/nature18318}{Nature {\bf 534}, 516-519 (2016)}.

\bibitem{TC} 
J. Zhang, P. W. Hess, A. Kyprianidis, P. Becker, A. Lee, J. Smith, G. Pagano, I.-D. Potirniche, A.C. Potter, A. Vishwanath, N. Y. Yao, and C. Monroe, 
{\it Observation of a discrete time crystal},
\href{https://www.nature.com/articles/nature21413}{Nature {\bf 543}, 217 (2017)}.

\bibitem{Kyprianidis2021}
A. Kyprianidis, F. Machado, W. Morong, P. Becker, K. S. Collins, D. V. Else, L. Feng, P. W. Hess, C. Nayak, G. Pagano, N. Y. Yao, and C. Monroe, 
{\it Observation of a prethermal discrete time crystal},
\href{https://www.science.org/doi/10.1126/science.abg8102}{Science {\bf 372}, 1192 (2021)}.

\bibitem{MBLi} 
J. Smith, A. Lee, P. Richerme, B. Neyenhuis, P. W. Hess, P. Hauke, M. Heyl, D. A. Huse, and C. Monroe, 
{\it Many-body localization in a quantum simulator with programmable random disorder},
\href{https://www.nature.com/articles/nphys3783}{Nat. Phys. {\bf 12}, 907 (2016)}.

\bibitem{morong2021observation}
W. Morong, F. Liu, P. Becker, K. S. Collins, L. Feng, A. Kyprianidis, G. Pagano, T. You, A. V. Gorshkov, and C. Monroe,
{\it Observation of Stark many-body localization without disorder},
\href{https://arxiv.org/abs/2102.07250}{arXiv:2102.07250}.

\bibitem{Kokail2019}
C. Kokail, C. Maier, R. van Bijnen, T. Brydges, M. K. Joshi, P. Jurcevic, C. A. Muschik, P. Silvi, R. Blatt, C. F. Roos, and P. Zoller, 
{\it Self-verifying variational quantum simulation of lattice models},
\href{https://www.nature.com/articles/s41586-019-1177-4}{Nature {\bf 569}, 355 (2019)}.

\bibitem{DPTm} 
J. Zhang, G. Pagano, P. W. Hess, A. Kyprianidis, P. Becker, H. B. Kaplan, A. V. Gorshkov, Z.-X. Gong, and C. Monroe, 
{\it Observation of a many-body dynamical phase transition with a 53-qubit quantum simulator},
\href{https://www.nature.com/articles/nature24654}{Nature {\bf 551}, 601 (2017)}.

\bibitem{DPTb}
P. Jurcevic, H. Shen, P. Hauke, C. Maier, T. Brydges, C. Hempel, B. P. Lanyon, M. Heyl, R. Blatt, and C. F. Roos,
{\it Direct Observation of Dynamical Quantum Phase Transitions in an Interacting Many-Body System},
\href{https://journals.aps.org/prl/abstract/10.1103/PhysRevLett.119.080501}{Phys. Rev. Lett. {\bf 119}, 080501 (2017)}.

\bibitem{Es} 
P. Jurcevic, B. P. Lanyon, P. Hauke, C. Hempel, P. Zoller, R. Blatt, and C. F. Roos,
{\it Quasiparticle engineering and entanglement propagation in a quantum many-body system},
\href{https://www.nature.com/articles/nature13461}{Nature {\bf 511}, 202 (2014)}.

\bibitem{richerme2014}
P. Richerme, Z.-X. Gong, A. Lee, C. Senko, J. Smith, M. Foss Feigh, S. Michalakis, A. V. Gorshkov, and C. Monroe, 
{\it Non-local propagation of correlations in quantum systems with long-range interactions},
\href{https://www.nature.com/articles/nature13450}{Nature {\bf 511}, 198 (2014)}.

\bibitem{tan2021domain}
W. L. Tan, P. Becker, F. Liu, G. Pagano, K. S. Collins, A. De, L. Feng, H. B. Kaplan, A. Kyprianidis, R. Lundgren, W. Morong, S. Whitsitt, A. V. Gorshkov, and C. Monroe, 
{\it Domain-wall confinement and dynamics in a quantum simulator},
\href{https://www.nature.com/articles/s41567-021-01194-3}{Nature Physics {\bf 6}, 742 (2021)}.

\bibitem{Kaplan2020}
H. B. Kaplan, L. Guo, W. L. Tan, A. De, F. Marquardt, G. Pagano, and C. Monroe, 
{\it Many-Body Dephasing in a Trapped-Ion Quantum Simulator}, 
\href{https://journals.aps.org/prl/abstract/10.1103/PhysRevLett.125.120605}{Phys. Rev. Lett. {\bf 125}, 120605 (2020)}.

\bibitem{rey}
A. Chu, J. Will, J. Arlt, C. Klempt, and A. M. Rey,
{\it Simulation of XXZ Spin Models using Sideband Transitions in Trapped Bosonic Gases},
\href{https://journals.aps.org/prl/abstract/10.1103/PhysRevLett.125.240504}{Phys. Rev. Lett. {\bf 125}, 240504 (2020)}.

\bibitem{geyer}
S. Geier, N. Thaicharoen, C. Hainaut, T. Franz, A. Salzinger, A. Tebben, D. Grimshandl, G. Z\"urn, and Matthias Weidem\"uller,
{\it Floquet Hamiltonian Engineering of an Isolated Many-Body Spin System},
\href{https://arxiv.org/abs/2105.01597}{arXiv:2105.01597}.

\bibitem{scholl}
P. Scholl, H. J. Williams, G. Bornet, F. Wallner, D. Barredo, T. Lahaye, A. Browaeys, L. Henriet, A. Signoles, C. Hainaut, T. Franz, S. Geier, A. Tebben, A. Salzinger, G. Z\"urn, M. Weidem\"uller,
{\it Microwave-engineering of programmable XXZ Hamiltonians in arrays of Rydberg atoms},
\href{https://arxiv.org/abs/2107.14459}{arXiv:2107.14459}.

\bibitem{molecules}
A. V. Gorshkov, S. R. Manmana, G. Chen, J. Ye, E. Demler, M. D. Lukin, and A. M. Rey,
{\it Tunable Superfluidity and Quantum Magnetism with Ultracold Polar Molecules},
\href{https://journals.aps.org/prl/abstract/10.1103/PhysRevLett.107.115301}{Phys. Rev. Lett. {\bf 107}, 115301 (2011).}

\bibitem{frerot}
Ir\'en\'ee Fr\'erot, Piero Naldesi, Tommaso Roscilde 
{\it Entanglement and fluctuations in the XXZ model with power-law interactions},
\href{https://journals.aps.org/prb/abstract/10.1103/PhysRevB.95.245111}{Phys. Rev. B {\bf 95}, 245111 (2017).}

\bibitem{mermin} 
N. D. Mermin and H. Wagner, 
{\it Absence of Ferromagnetism or Antiferromagnetism in One- or Two-Dimensional Isotropic Heisenberg Models}, 
\href{https://journals.aps.org/prl/abstract/10.1103/PhysRevLett.17.1133}{Phys. Rev. Lett. {\bf 17}, 1133 (1966)}.

\bibitem{abram}
M. Abramowitz and I. A. Stegun 
{\it  Handbook of Mathematical Functions}.
 (New York: Dover) 1964. 

\bibitem{grad}
I. S. Gradshteyn and I. M. Ryzhik,
{\it  Tables of Integrals, Series, and Products}
(Amsterdam: Elsevier) 2007.

\bibitem{olver}
F. W. J. Olver, D. W. Lozier, R. F.  Boisvert and C. W. Clark,
{\it NIST Handbook of Mathematical Functions},
(Cambridge: Cambridge University Press) 2010.

\bibitem{dmrg} 
S. R. White and A. E. Feiguin, 
{\it Real-Time Evolution Using the Density Matrix Renormalization Group},
\href{https://journals.aps.org/prl/abstract/10.1103/PhysRevLett.93.076401}{
Phys. Rev. Lett. {\bf 93}, 076401 (2004)}. 

\bibitem{dmrg2}
A. E. Feiguin and S. R. White,
{\it Time-step targeting methods for real-time dynamics using the density matrix renormalization group},
\href{https://journals.aps.org/prb/abstract/10.1103/PhysRevB.72.020404}{Phys. Rev. B {\bf 72}, 020404(R) (2005)}.

\bibitem{piazza}
F. Mivehvar, H. Ritsch, and F. Piazza,
{\it Cavity-quantum-electrodynamical toolbox for quantum magnetism},
\href{https://journals.aps.org/prl/abstract/10.1103/PhysRevLett.122.113603}{Phys. Rev. Lett. {\bf 122}, 113603 (2019)}.

\bibitem{Bermudez2017}
A. Bermudez, L. Tagliacozzo, G. Sierra, and P. Richerme, 
{\it Long-range Heisenberg models in quasiperiodically driven crystals of trapped ions}.
\href{https://journals.aps.org/prb/abstract/10.1103/PhysRevB.95.024431}{Phys. Rev. B {\bf 95}, 024431 (2017)}.

\bibitem{porras2004}
D. Porras, and I. Cirac, 
{\it Effective quantum spin systems with trapped ions}, 
\href{https://journals.aps.org/prl/abstract/10.1103/PhysRevLett.92.207901}{Phys. Rev. Lett. {\bf 92}, 207901 (2004)}.

\bibitem{deng}  
X.-L. Deng, D. Porras,  and J. I. Cirac, 
{\it Effective spin quantum phases in systems of trapped ions},
\href{https://journals.aps.org/pra/abstract/10.1103/PhysRevA.72.063407}{Phys. Rev. A \textbf{72}, 063407 (2005)}.

\bibitem{Davoudi2020}
Z. Davoudi,  M. Hafezi, C. Monroe,  G. Pagano,  A. Seif, and A. Shaw, 
{\it Towards analog quantum simulations of lattice gauge theories with trapped ions}, 
\href{https://journals.aps.org/prresearch/abstract/10.1103/PhysRevResearch.2.023015}{Phys. Rev. Res. {\bf 2}, 023015 (2020)}.

\bibitem{Leibfried2003}
D. Leibfried, B. DeMarco, V. Meyer, D. Lucas, M. Barrett, J. Britton, W. M. Itano, B. Jelenković, C. Langer, T. Rosenband, and D. J. Wineland, 
{\it Experimental demonstration of a robust, high-fidelity geometric two ion-qubit phase gate}, 
\href{https://www.nature.com/articles/nature01492}{Nature {\bf 422}, 412 (2003)}.

\bibitem{Lloyd1996}
S. Lloyd, 
{\it Universal Quantum Simulators}, 
\href{https://www.science.org/doi/abs/10.1126/science.273.5278.1073}{Science {\bf 273}, 1073, (1996)}.

\bibitem{Lanyon2011}
B. P. Lanyon, C. Hempel, D. Nigg, M. Müller, R. Gerritsma, F. Zähringer, P. Schindler, J. T. Barreiro, M. Rambach, G. Kirchmair, M. Hennrich, P. Zoller, R. Blatt, and C. F. Roos,
 {\it Universal Digital Quantum Simulation with Trapped Ions}, 
 \href{https://www.science.org/doi/abs/10.1126/science.1208001}{Science {\bf 334}, 57 (2011)}.

\bibitem{Heyl2019}
M. Heyl, P. Hauke, and P. Zoller, 
{\it Quantum localization bounds Trotter errors in digital quantum simulation}, 
\href{https://www.science.org/doi/10.1126/sciadv.aau8342}{Sc. Adv. {\bf 5}  8342, (2019)}. 

\bibitem{haldane}
F. D. M. Haldane, 
{\it Nonlinear Field Theory of Large-Spin Heisenberg Antiferromagnets: Semiclassically Quantized Solitons of the One-Dimensional Easy-Axis Néel State}
\href{https://journals.aps.org/prl/abstract/10.1103/PhysRevLett.50.1153}{Phys. Rev. Lett. {\bf 50}, 1153 (1983)}.

\bibitem{gong}
Z.-X. Gong, M. F. Maghrebi, A. Hu, M. L. Wall, M. Foss-Feig, and A. V. Gorshkov,
{\it Topological phases with long-range interactions}
\href{https://journals.aps.org/prb/abstract/10.1103/PhysRevB.93.041102}{Phys. Rev. B {\bf 93}, 041102 (2016)}.

\bibitem{bluvstein}
D. Bluvstein, A. Omran, H. Levine, A. Keesling, G. Semeghini, S. Ebadi, T. T. Wang, A. A. Michailidis, N. Maskara, W. W. Ho, S. Choi, M. Serbyn, M. Greiner, V. Vuletic, and M. D. Lukin,
{\it Controlling many-body dynamics with driven quantum scars in Rydberg atom arrays},
\href{https://science.sciencemag.org/content/371/6536/1355}{Science {\bf 371}, 1355-1359 (2021)}.

\bibitem{celi}
A. Celi, B. Vermersch, O. Viyuela, H. Pichler, M. D. Lukin, and P. Zoller,
{\it Emerging 2D Gauge theories in Rydberg configurable arrays},
\href{https://journals.aps.org/prx/abstract/10.1103/PhysRevX.10.021057}{Phys. Rev. X {10}, 021057 (2020)}.

\bibitem{borla}
U. Borla, R. Verresen, F. Grusdt, and S. Moroz,
{\it Confined phases of one-dimensional spinless fermions coupled to Z2 gauge theory},
\href{https://journals.aps.org/prl/abstract/10.1103/PhysRevLett.124.120503}{Phys. Rev. Lett. {\bf 124}, 120503 (2020)}.

\bibitem{kerbic}
M. Kebrič, L. Barbiero, C. Reinmoser, U. Schollwöck, and F. Grusdt,
{\it Confinement and Mott transitions of dynamical charges in 1D lattice gauge theories},
\href{https://arxiv.org/abs/2102.08375}{arXiv:2102.08375}.



\end{thebibliography}
\end{document}